\documentclass[journal]{IEEEtranTIE}
\usepackage{graphicx}
\usepackage{cite}
\usepackage{picinpar}
\usepackage{amsmath}
\usepackage{centernot}
\usepackage{stfloats}
\usepackage{url}
\usepackage{flushend}
\usepackage[latin1]{inputenc}
\usepackage{colortbl}
\usepackage{soul}
\usepackage{multirow}
\usepackage{pifont}
\usepackage{color}
\usepackage{alltt}
\usepackage[hidelinks]{hyperref}
\usepackage{enumerate}
\usepackage{siunitx}
\usepackage{breakurl}
\usepackage{epstopdf}
\usepackage{pbox}

\usepackage{import}
\usepackage{graphicx}
\usepackage{xcolor}
\usepackage{pgfplots}
\usepackage{tikz}
\usepgfplotslibrary{colorbrewer}
\usepackage{amsfonts}
\usepackage{amssymb}
\usepackage{xfrac}
\usepackage[compatibility=false]{caption}
\usepackage[hypcap,labelformat=simple]{subcaption}
\usepackage{float}

\newcommand{\vect}[1]{\mathbf{#1}} 

\begin{document}
\title{Real-Time Refocusing using an FPGA-based Standard Plenoptic Camera}

\author{
	\vskip 1em
	{
	Christopher Hahne,
	Andrew Lumsdaine,~\IEEEmembership{Senior Member,~IEEE},
	Amar Aggoun, \\ and
	Vladan Velisavljevic,~\IEEEmembership{Senior Member,~IEEE}
	}

	\thanks{		
		{
		Manuscript received June 11, 2017; revised January 9, 2018; accepted March 1, 2018.
		This work was supported in part by the EU's 7th Framework Programme under Grant EU-FP7 ICT-2010-248420.
		
		Christopher Hahne is with trinamiX GmbH (BASF), Ludwigshafen, Germany (e-mail: info@christopherhahne.de). 
		
		Andrew Lumsdaine is with the Northwest Institute for Advanced Computing, University of Washington, Seattle, WA 98195 USA (e-mail: al75@uw.edu). 
		
		Amar Aggoun is with the School of Mathematics and Computer Science, University of Wolverhampton, Wolverhampton, UK (e-mail: a.aggoun@wlv.ac.uk).
		
		Vladan Velisavljevic is with the School of Computer Science and Technology, Luton, UK (vladan.velisavljevic@beds.ac.uk).	
	
		}
	}
}

\maketitle
	
\begin{abstract}
%
Plenoptic cameras are receiving increasing attention in scientific and commercial applications because they capture the entire structure of light in a scene, enabling optical transforms (such as focusing) to be applied computationally after the fact, rather than once and for all at the time a picture is taken.
%
In many settings, real-time interactive performance is also desired, which in turn requires significant computational power due to the large amount of data required to represent a plenoptic image.
Although GPUs have been shown to provide acceptable performance for real-time plenoptic rendering, their cost and power requirements make them prohibitive for embedded uses (such as in-camera).
%
On the other hand, the computation to accomplish plenoptic rendering is well-structured, suggesting the use of specialized hardware.
Accordingly, this paper presents an array of {switch-driven} Finite Impulse Response (FIR) filters, implemented with FPGA to accomplish high-throughput spatial-domain rendering.
%
The proposed architecture provides a {power-efficient} rendering hardware design suitable for {full-video} applications as required in broadcasting or 
cinematography. %
%
%
A benchmark assessment of the proposed hardware implementation shows that real-time performance can readily be achieved, with a one order of magnitude performance improvement over a GPU implementation and three orders of magnitude performance improvement over a general-purpose CPU implementation.
\end{abstract}


\markboth{IEEE TRANSACTIONS ON INDUSTRIAL ELECTRONICS}%
{}

\definecolor{limegreen}{rgb}{0.2, 0.8, 0.2}
\definecolor{forestgreen}{rgb}{0.13, 0.55, 0.13}
\definecolor{greenhtml}{rgb}{0.0, 0.5, 0.0}

\section{Introduction}
%
Over the last two decades, several studies have reported methods to computationally
render
varyingly focused images from a single lightfield photograph\if taken by a plenoptic camera\fi~\cite{Isaksen:2000:DRL:344779.344929,NGLEV,NGACM,LUMSFULL,RAYTRIX,Hahne:14:SPIE,Perez:2014,Hahne:OPEX:16}. 
In addition to spatial information, 
lightfields contain directional information, acquired by capturing an array of two-dimensional \mbox{(2-D)} spatial images with either multiple conventional cameras~\cite{LEVHAN,Isaksen:2000:DRL:344779.344929,Yang:2002,VENKATARAMAN} or by attaching a \textit{Micro Lens Array} (MLA) to a single image recording device~\cite{Lippmann:1908,AW,NGLEV}. 
%
In science, lightfield cameras are also known as plenoptic cameras derived from the Latin and Greek roots meaning 'full view'~\cite{AB,AW}. %
For industrial applications, MLAs are preferred to simple pinholes or coded-aperture patterns due to improved light-gather capability and to multi-aperture systems due to compact form-factor.
%
%
A study carried out by Ng~{\it et al.}~\cite{NG} has found that the maximum directional information is recorded when placing the micro lenses one focal length away from the image sensor. However, a follow-up study re-investigated this and showed that it is possible to flexibly trade off directional and spatial resolution by shifting the MLA with respect to the sensor~\cite{LUMSFOC,LUMSFULL}. %
In this paper, we refer to the former design as the \textit{Standard Plenoptic Camera} (SPC) and the latter as the \textit{Focused Plenoptic Camera} (FPC).
%
While researchers have developed a number of approaches to plenoptic camera design~\cite{Veeraraghavan:2007:DPM:1276377.1276463,Xu:12}, the  rendering (or focusing) process remains computationally intensive, posing a core challenge to the computer vision field. 
%
%

One motivating industrial performance-sensitive application for plenoptic cameras is in cinematography, where the use of plenoptic source video can greatly enhance the flexibility and creativity in capture and production. %
For example, since the optical parameters are not irrevocably set at the time the video is captured, focus or depth of field can easily be adjusted in post-production.  Moreover, new creative effects can be applied, including non-physical optical effects. Plenoptic video can also be used to create stereo pairs for 3D viewing -- with the important advantage over stereo capture that different videos can be created for different devices, each having parallax suited for the particular device~\cite{Hahne:18:IJCV}. Finally, 2D and 3D production can use significantly different effects for directing the viewer's attention (depth of field is not as useful in 3D as 2D, for example).  With plenoptic source video, 2D and 3D can be rendered from the same source, with different creative effects for each. 
We note that Lytro, one of the earliest manufacturers of plenoptic cameras,
has recently announced a video lightfield camera to the broadcast and cinematography market~\cite{LytroCinemaPressRelease:16}.
In any of these scenarios, high rendering performance is essential.  For preview and for post-production, rendering of each video frame must be accomplished at the video frame rate, regardless of the effects and adjustments being applied. %

An early attempt at high-performance rendering was based on the projection slice theorem, which rendered images with lower-dimensional slices of the lightfield in the Fourier domain~\cite{NGACM,Mhabary:16:SPIE}. This procedure is also known as \textit{Fourier Slice Photography} (FSP). 
%
%
Although FSP has the potential to be efficient when rendering a large number of focused images from the same lightfield, there are significant overheads in this approach that limit its practical application.  Real-time rendering in the spatial-domain has been achieved with \textit{Graphics Processing Units} (GPUs)~\cite{lumsdaine12:_plenoptic_gpus}, but the cost and power associated with GPUs make their use in embedded settings (for example) impractical.
Accordingly, it is the goal of this study to devise and demonstrate a special-purpose hardware architecture that performs real-time rendering in the spatial-domain based on serially incoming video frames. 
We propose an array of semi-systolic \textit{Finite Impulse Response} (FIR) filters designed for high data throughput. 
Moreover, we realize the rendering convolution kernel in FIR fashion by introducing 
switches to the filter distribution network. 
For power efficiency and configuration flexibility, the proposed design is implemented with a \textit{Field Programmable Gate Array}~(FPGA). 
As distinguished from previous studies, our hardware design accomplishes a computation time of less than 100~$\mu$s for a single refocused frame with 3201-by-3201 pixel resolution when running at 100 MHz pixel clock frequency. This outperforms earlier studies in the field, which we further demonstrate with benchmarks against a GPU and a CPU Matlab implementation. \par%
%
%
%
%
The organization of this paper is as follows. Section~\ref{sec:background} presents recent developments in the field of FSP and SPC lightfield modeling to serve as a starting point for refocusing in spatial-domain. 
Section~\ref{sec:Shift} imposes requirements on the filter module architecture and presents a solution based on {switch-driven} FIR filters. 
The proposed hardware design is examined in Section~\ref{sec:fpgaimplementation}, using a \textit{Hardware Description Language} (HDL) for FPGAs (see Appendix) and by benchmarks with an alternative GPU-based implementation. 
Conclusions and suggestions for further work are presented in 
Section~\ref{sec:discussion}.
\section{Related Work}
\label{sec:background}
%
\subsection{Background}
A lightfield can be retrieved by light rays intersecting two consecutively-placed two-dimensional \mbox{(2-D)} planes of known relative position~\cite{LEVHAN}. Intersections of a single ray at two \mbox{2-D} planes yield 4 coordinates in total, thus making up a four-dimensional (4-D) light ray parametrization. Due to its simplicity, this conceptual model has gained popularity among scientists in the field of computer vision. A related one-plane parameterization based on position and angle
can also be used~\cite{LUMSFULL,LUMSFOC}.
In the celebrated work by Ng~{\it et al.}~\cite{NGACM} , a raw captured 4-D lightfield is transformed to the Fourier domain to achieve refocusing using the projection-slice theorem. %
%
Unfortunately, the process of taking Fourier transforms, interpolating for slicing, and then taking inverse transforms introduces significant computational overhead, making FSP unsuitable for real-time rendering.
%
This assumption was confirmed by Mhabary~{\it et al.}~\cite{Mhabary:16:SPIE}, who have worked to advance FSP by employing a fractional Fourier transform. However, the authors conclude that the integral projection operator in the spatial-domain is faster when computing only a single refocused image from a lightfield. 
The suitability of refocusing in the spatial-domain was further confirmed by 
Lumsdaine~{\it et al.}, who demonstrated real-time rendering performance using GPU hardware~\cite{lumsdaine12:_plenoptic_gpus}.
For these reasons, our approach in this paper is based on rendering in the spatial-domain. \par
The main concept of computation time improvements using FPGAs builds on the principle of parallelization and pipelining~\cite{BAILEY}. A pipeline comprises chained processor blocks fed with serialized data that are processed sequentially. 
Speed up is obtained by processing data chunks in one processor unit while subsequent data chunks are handled in preceding units. 
Hence, the benefit of pipelining is that serialized data chunks are processed at the same time while processor units perform different tasks. While data serialization limits a specific task to be computed with one single operation at a time, e.g., one pixel after another, parallelized data streams allow a computing system to perform at least two operations of the same type simultaneously. Parallelization can be thought of as duplicating processor pipelines, which requires synchronized parallel data streams as input signals. Letting the degree of parallelization be~$\iota$, the computation time in image processing may be minimized to $\mathcal{O}\left(\sfrac{K^2}{\iota}\right)$ if \mbox{2-D} image dimensions consist of $K$ samples each and provided that both computation systems run at the same clock frequency. Consequently, the one-dimensional \mbox{(1-D)} parallelization limit is reached where $\iota=L$ for image rows and $\iota=K$ for image columns, which is the ideal scenario in terms of parallelizing data processes. \par%
Early work in the field of embedded plenoptic imaging was reported by Rodr\'iguez-Ramos~{\it et al.}~\cite{Rodriguez:2010}, who employed an FPGA to process plenoptic data with the aim of analyzing wavefront measurements. %
Another interesting approach, reported by Wimalagunarathne~{\it et al.}~\cite{Randeel:2012}, proposed a design to render computationally focused photographs from a set of multi-view images using \textit{Infinite Impulse Response} (IIR) filters. %
Work on real-time rendering from FPC captures was presented in~\cite{lumsdaine12:_plenoptic_gpus}. %
The first reported hardware design for performing {real-time} rendering from SPC captures was presented by Hahne~{et al.}~\cite{Hahne:14:SPIE}. Shortly thereafter, P\'{e}rez~{\it et al.}~\cite{Perez:2014} published an article addressing the same topic. %
The authors demonstrated significant computation time improvements compared to run times based on a \textit{Central Processing Unit} (CPU) system that was programmed using an object-oriented language. A theoretical comparison of our method with that of P\'{e}rez~{\it et al.}~\cite{Perez:2014} is carried out at the end of Section~\ref{sec:Shift}. 
%
%
\subsection{SPC Ray Model}
\label{ssec:rayModel}
\begin{figure}[ht]
	\centering
	\includegraphics[width=.75\linewidth]{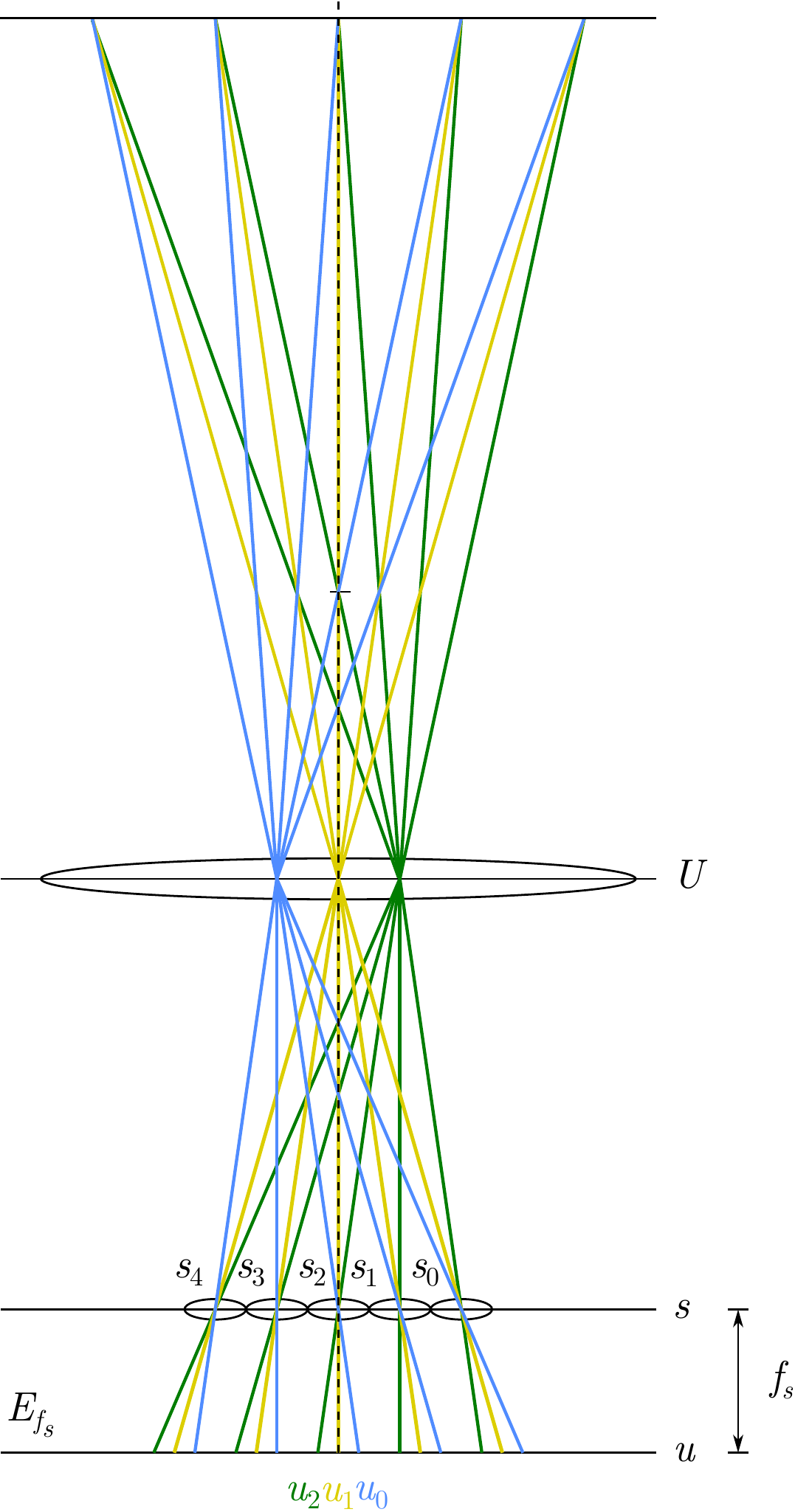}
	\caption[]{SPC ray model (borrowed from \cite{Hahne:OPEX:16}) with micro lens chief rays traveling through the MLA plane $s$ and main lens plane $U$, which is depicted as a thin lens. Lightfield intensities captured at the sensor plane are denoted as $E_{f_s}\left[s_j \, , \, u_{c+i}\right]$ for the 1-D case. Chief ray colors in a micro image indicate angular samples $u_{c+i}$.
	\label{fig:refinedModel}}
\end{figure}

Development of a computationally efficient refocusing algorithm requires knowledge about the ray geometrical properties in a plenoptic camera. To conceive a refocusing hardware architecture in spatial-domain, we employ a ray model reported by Hahne~\textit{et al.}~\cite{Hahne:OPEX:16}, which is based on paraxial optics. %
The model is depicted in Fig.~\ref{fig:refinedModel} and builds on the assumption that image sensor plane and MLA are separated by one focal length, $f_s$, such that the MLA is focused to infinity, which is in accordance with Ng's concept of a plenoptic camera~\cite{NG}. 
To understand lightfield imaging in an SPC, as in the Lytro setup~\cite{LytroCinemaPressRelease:16}, one may regard a main lens image of an object plane to be focused on the MLA plane. In this case, 
the focused light rays converge to the micro lens and diverge when leaving it to form a micro image (see Fig.~\ref{fig:refinedModel}). A pixelated light-sensitive detector placed behind the MLA captures angular portions of the incident-divergent beam. Each angular sample in this micro image corresponds to the same focused spatial point in space observed from different views. This point's intensity is recovered when integrating all micro image samples. \par
%
%
%
We denote a lightfield captured by an SPC in the following way. For clarity, only the horizontal cross-section is regarded hereafter. 
In the angular domain $u$, we start counting samples from \textit{Micro Image Centers} (MICs), which serve as a reference positions $c = \sfrac{(M - 1)}{2}$ where $M$ denotes a consistent total number of samples for each micro image in one dimension. Micro images are seen to be radially symmetric and horizontally indexed by $c+i$, with $i \in [-c \, .. \, c]$. Horizontal lightfield positions are then given as $\left[s_j \, , \, u_{c+i}\right]$ with $j$ as the \mbox{1-D} index of a respective micro lens $s_j$. 
All micro images together form a light field image with its cross-sectional representation $E_{f_s}\left[s_j \, , \, u_{c+i}\right]$ where $E_{f_s}$ denotes a pixel's illuminance. 
%
As demonstrated in~\cite{Hahne:OPEX:16}, a horizontal \mbox{cross-section} of a lightfield image can be refocused by employing
\begin{align}
E'_{a}\left[s_j\right] &= \sum_{i=-c}^{c} \, \frac{1}{M} \, E_{f_s}\left[ s_{j+a(c-i)} \, , \, u_{c+i}\right]  \, \, , \quad  a \in \mathbb{Q} \label{eq:refocusingAverage}
\end{align}
where $a$ adjusts the synthetic focus. Equation~\eqref{eq:refocusingAverage} can also be applied to the vertical dimension. \par%
%
Since images acquired by an SPC do not feature the $E_{f_s}\left[s_j \, , \, u_{c+i}\right]$ notation, it is convenient to define an index translation formula considering the lightfield photograph to be of two regular sensor dimensions $\left[x_k \, , \, y_l\right]$ as if taken by a conventional sensor. Indices are then converted by
\begin{align}
k &=j \times M+c+i \label{eq:translate}
\end{align}
in the horizontal dimension meaning that $\left[x_k\right]$ is formed by 
$\left[x_{j \times M+c+i}\right]$ to replace $\left[s_j \, , \, u_{c+i}\right] \label{eq:translate2}$.
%
This concept of index translation may be similarly extended to the vertical domain. 
%
\section{Filter Design}
\label{sec:Shift}
\begin{figure}[ht]
	\includegraphics[width=\linewidth]{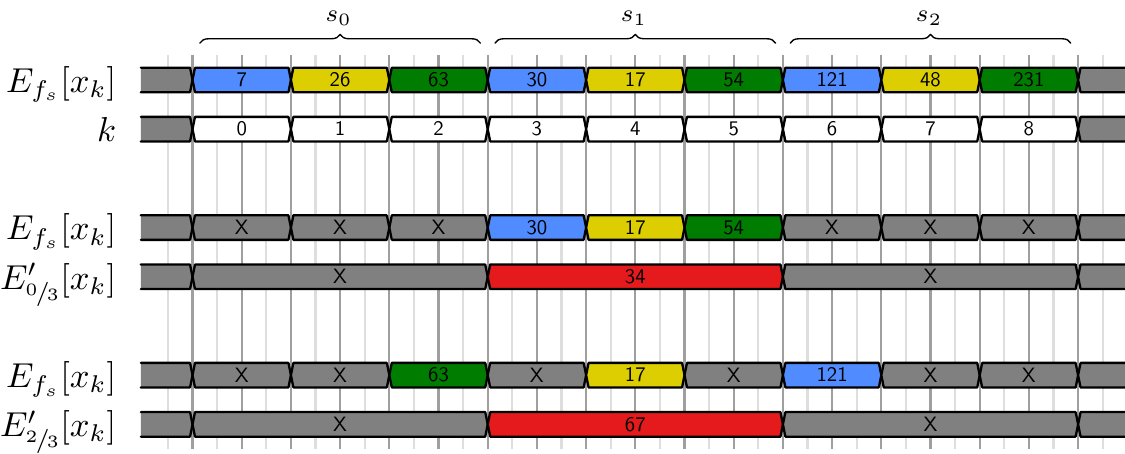}
	\caption[Processing requirements for the hardware architecture]{Processing requirements for the hardware architecture. The diagram shows exemplary input illuminance values $E_{f_s}$ (see Fig.~\ref{fig:refinedModel}) subdivided into micro images $s_j$ and synthesized output values $E'_a$ at a desired refocused image plane $a$.}
	\label{fig:flowDiagramProblem}
\end{figure}

An efficient hardware design that enables an FPGA to refocus in {real-time} may be conceptualized on the basis of the lightfield ray model presented in Section~\ref{sec:background}. %
The upper data line of Fig.~\ref{fig:flowDiagramProblem} depicts discrete and quantized illuminance values $E_{f_s}\left[x_k\right]$ of a single horizontal row that is part of a calibrated lightfield image. Lightfield calibration implies MIC detection and rendering procedures to obtain a consistent micro image size ($M$). The computational refocusing synthesis given in Section~\ref{sec:background} reveals that pixels involved in the integration process expose interleaved neighborhood relations, which exclusively depend on $a$. This phenomenon is illustrated by the data flow diagram in Fig.~\ref{fig:flowDiagramProblem}, where respective pixels are highlighted for two exemplary refocusing settings: $a=\sfrac{0}{3}$ and $a=\sfrac{2}{3}$. Here, each color corresponds to a chief ray in the model in Fig.~\ref{fig:refinedModel}, with $M=3$ where yellow represents the MIC pixel. In this section, a hardware architecture is devised that accomplishes signal processing according to Eq.~(\ref{eq:refocusingAverage}) as depicted in Fig.~\ref{fig:flowDiagramProblem}. \par%
\noindent
%
On the supposition that a horizontal cross-section of a captured lightfield $E_{f_s}\left[x_k\right]$ is a linear, time-invariant system, the integral projection in Eq.~(\ref{eq:refocusingAverage}) may be represented as a discrete FIR convolution formula. 
Following the $\left[s_j,u_{c+i}\right]$ to $\left[x_k\right]$ translation in Section~\ref{sec:background}, \mbox{1-D} refocusing can be given by
\begin{align}
E'_{a}\left[x_{k}\right] &=\sum_{i=0}^{M-1} \frac{1}{M} \, E_{f_s} \left[x_{k'+i(aM-1)}\right]  \, \, , \quad  a \in \mathbb{Z}
\label{eq:refocusingAverageX}
\end{align}
with
\begin{align}
k'=(k+1)\times M-1
\end{align}
taking care of a correct integral projection, which inevitably reduces the number of samples in the rendered output image. Equation~(\ref{eq:refocusingAverageX}) aims at complying with the classical FIR filter notation, however with indices in subscripts for consistency reasons and to let $x$ signify the domain and coordinate direction. Upon closer examination, one may notice that the impulse response is represented by a constant coefficient $\sfrac{1}{M}$, which is a consequence of weighting pixels equally during the integration process. Note that $i \in \left[0 \, ..  \, M-1\right]$ in the following. \par%
In contrast to Eq.~(\ref{eq:refocusingAverageX}), we seek to reproduce an output image with a resolution numerically equal to that of the raw sensor image. To compensate for sample reduction in the integral projection process, the overall sensor resolution may be retained by upsampling the spatial-domain during image formation. Besides, it will be shown hereafter that our proposed upsampling scheme enables interpolation of refocused depth planes. \par%
To break down the complexity, we devise one filtering function per refocusing slice $a$ that 
qualifies for FIR filter implementation. Regardless of the micro image resolution $M$, a filter that computes a refocusing slice with $a=0$ in horizontal direction reads
\begin{align}
E'_{\sfrac{0}{M}}\left[x_k\right] &= \sum_{i=0}^{M-1} \frac{1}{M} \, E_{f_s}\left[x_{k-i-\bmod{(k+1, \, M)}}\right] \, , \label{eq:refocusingX:0}
\end{align}
when $k\in\{0, ..., K-1\}$. Term $\bmod{(k+1, \, M)}$ comprises a \textit{Nearest-Neighbor} (NN) interpolation ensuring that the numerical output image resolution matches that of the input. A synthetically focused image where $a=1$ is formed by
\begin{align}
E'_{\sfrac{M}{M}}\left[x_k\right] &= \sum_{i=0}^{M-1} \frac{1}{M} \, E_{f_s}\left[x_{k+i(m-1)}\right] \, . \label{eq:refocusingX:3}
\end{align}
Synthesis equations for different $a=\sfrac{a'}{M}$ are retrieved by reverse-engineering. Probably, the most straightforward refocusing filter kernel function is given by
\begin{align}
E'_{\sfrac{1}{M}}\left[x_k\right] &= \sum_{i=0}^{M-1} \frac{1}{M} \, E_{f_s}\left[x_{k-i}\right] \, , \label{eq:refocusingX:1}
\end{align}
which computes refocusing slice $a=\sfrac{1}{M}$. When implementing Eq.~\eqref{eq:refocusingX:1} as an FIR filter, it becomes obvious that the number of filter taps amounts to $M$. A VHDL implementation using this filter type with $M=5$ is provided in supplementary material. 
In the following, we demonstrate a refocusing hardware architecture that is adapted to an SPC with $M=3$. Then, a photograph refocused with $a=\sfrac{2}{3}$ is computed by
\begin{align}
E'_{\sfrac{2}{3}}\left[x_k\right] &= \sum_{i=0}^{3-1} \frac{1}{3} \, E_{f_s}\left[x_{k-i+|\lceil \bmod{(k+1, \, 3)}/3\rceil-1|\times(i-1)}\right] \, . \label{eq:refocusingX:2}
\end{align}
where $\lceil\cdot\rceil$ is the ceiling and $|\cdot|$ the absolute value operator. An exemplary step in the computation of $E'_{\sfrac{2}{3}}[x_k]$ would be
\begin{align}
E'_{\sfrac{2}{3}}\left[x_3\right] &= \frac{1}{3} E_{f_s}\left[x_3\right] + \frac{1}{3} E_{f_s}\left[x_2\right] + \frac{1}{3} E_{f_s}\left[x_1\right] \, . \label{eq:refocusingXexample:1}
\end{align}
Here, fractions $\sfrac{1}{3}$ can be regarded as multipliers, denoted as $h_0$, which are identical for each pixel such that \mbox{$h_0=\sfrac{1}{M}$}. On condition that incoming images are underexposed and clipping is prevented, it is noteworthy that multipliers are redundant and thus can be left out\if (see Section~\ref{sec:refocusing} for more details)\fi. 
\subsection{Semi-Systolic Modules}
\label{ssec:horShift}
Equations \eqref{eq:refocusingX:0}--\eqref{eq:refocusingX:2} are implemented with a systolic filter design. Systolic arrays broadcast input data to many \textit{Processing Elements} (PE)s. As shown, all wired connections in a systolic filter contain at least one latch driven by the same clock signal. 
\emph{Semi-systolic} designs omit these latches. All of the remaining designs that we consider are semi-systolic, but latches can be added for systolic FPGA implementation purposes.
Descriptive information about systolic arrangements can be found in~\cite{Systolic}.

\begin{figure}[ht]
	\centering
	\includegraphics[width=1.02\linewidth]{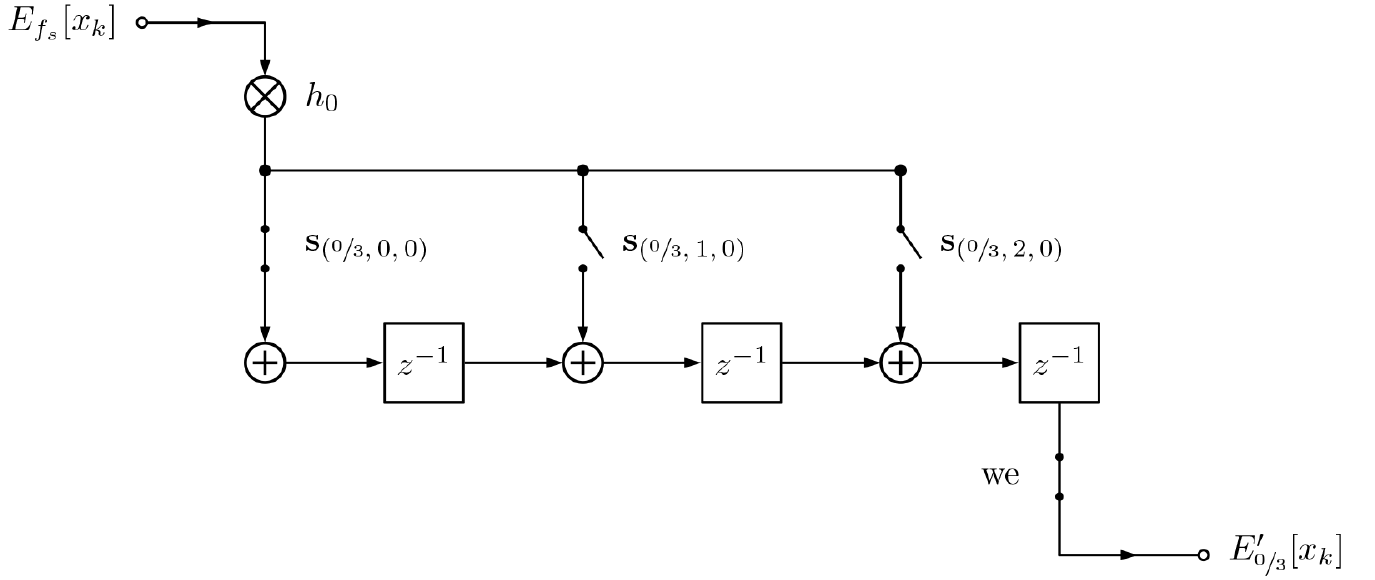}
	\caption[1-D semi-systolic FIR filter for sub-pixel shift $a =\sfrac{0}{3}$]{1-D semi-systolic FIR filter for sub-pixel shift $a =\sfrac{0}{3}$.\label{fig:FIR0}}
\end{figure}

A positive side effect of the systolic filter is that it can be exploited for an NN-interpolation in micro images. By letting the upsampling factor be the number of micro image samples $M$, the resolution loss in integral projection is compensated, since incoming and outgoing resolution are the same. Naturally, the interpolation method can be more sophisticated, which in turn requires intermediate calculations, causing delays and an increasing number of occupied logic gates. Closer inspection of Eq.~(\ref{eq:refocusingX:3}) reveals that pixels which need to be integrated are interlaced. Thereby, gaps between merged pixels grow with ascending $a$ and extend the filter length. The omission of pixels within gaps is realized with switches. A switch-controlled semi-systolic FIR filter design of Eq.~(\ref{eq:refocusingX:0}) with multiplier $h_0$ is depicted in Fig.~\ref{fig:FIR0}.
\noindent
\par
In this design, switch states are controlled by bits in a \mbox{2-D} vector field denoted as $\vect{s}_{(a,\,w,\,p)}$ which is given by 
\begin{align}
\vect{s}_{\left(\sfrac{0}{3},\,w,\,p\right)} =
\begin{bmatrix}
1 & 0 & 0 \\
0 & 1 & 0 \\
0 & 0 & 1 \\
\end{bmatrix}
\, .
\end{align}
if $a=\sfrac{0}{3}$. Depending on refocusing parameter $a$, switch state matrices $\vect{s}_{\left(a,\,w,\,p\right)}$ contain binary numbers with columns indexed by $w$ for the state of each switch in the FIR filter and with rows indexed by $p$, which loads a new row of switch states when incremented. In addition, a \textit{write enable} (we) switch helps to prevent intermediate falsified values from being streamed out.
\begin{figure}[ht]
	\includegraphics[width=\linewidth]{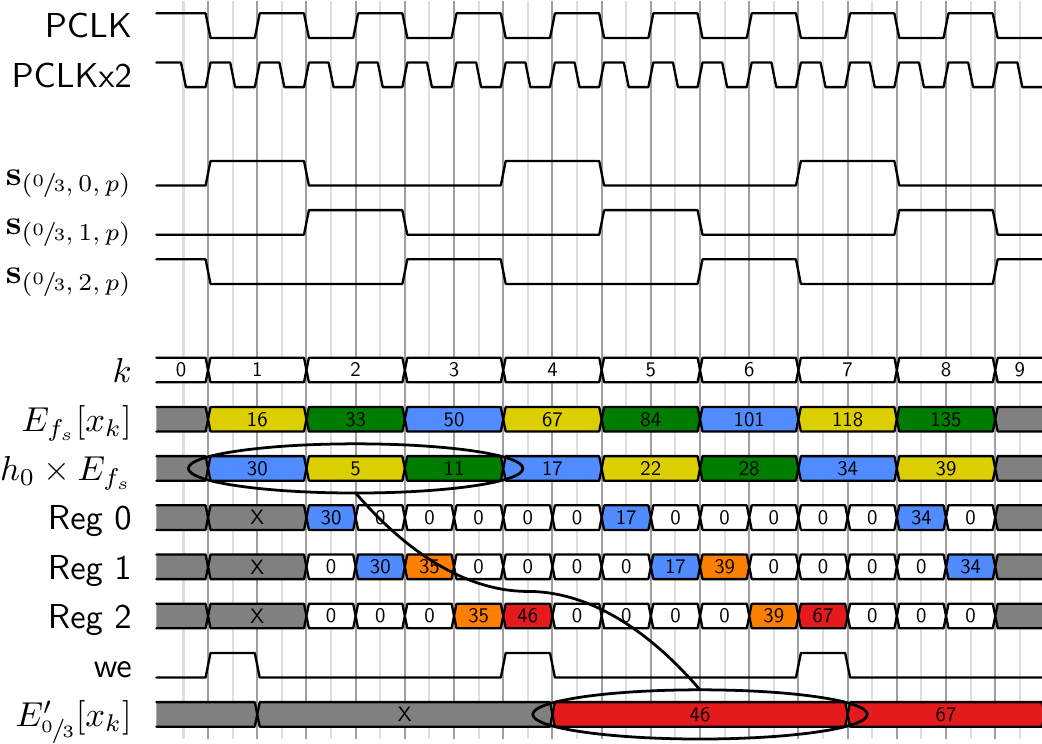}
	\caption[Timing diagram of 1-D semi-systolic FIR filter for $a=\sfrac{0}{3}$]{Timing diagram of FIR filter module with $a=\sfrac{0}{3}$.}
	\label{fig:flowDiagram0}
\end{figure}\noindent
%
For better comprehension, a timing diagram in Fig.~\ref{fig:flowDiagram0} visualizes the computational concept of the FIR design from Fig.~\ref{fig:FIR0}. Here, the pixel clock signal is given as PCLK. Furthermore, the proposed architecture employs the doubled pixel clock PCLKx2 with a time period $T_{\text{PCLKx2}}=T_{\text{PCLK}}/2$ to shift and add pixel values in a single pixel clock cycle $T_{\text{PCLK}}$. It is also seen that a new row of switch states is called by incrementing $p$ every pixel clock cycle. Numbers in the data streams represent unsigned decimal 8-bit gray-scale values, which are multiplied with $h_0=\sfrac{1}{3}$. Pixel colors match those of the SPC ray model in Fig.~\ref{fig:refinedModel} representing chief ray positions in micro images with $M=3$. Orange color highlights interim results and red signifies \mbox{1-D} refocused output data. Oval circles indicate that the sum of divided micro image pixels is reflected in the output pixel $E'_{\sfrac{0}{3}}\left[x_k\right]$. The filter includes an NN-interpolation upsampling the micro image resolution by factor~3. \par
\begin{figure}[ht]
	\centering
	\includegraphics[width=1.02\linewidth]{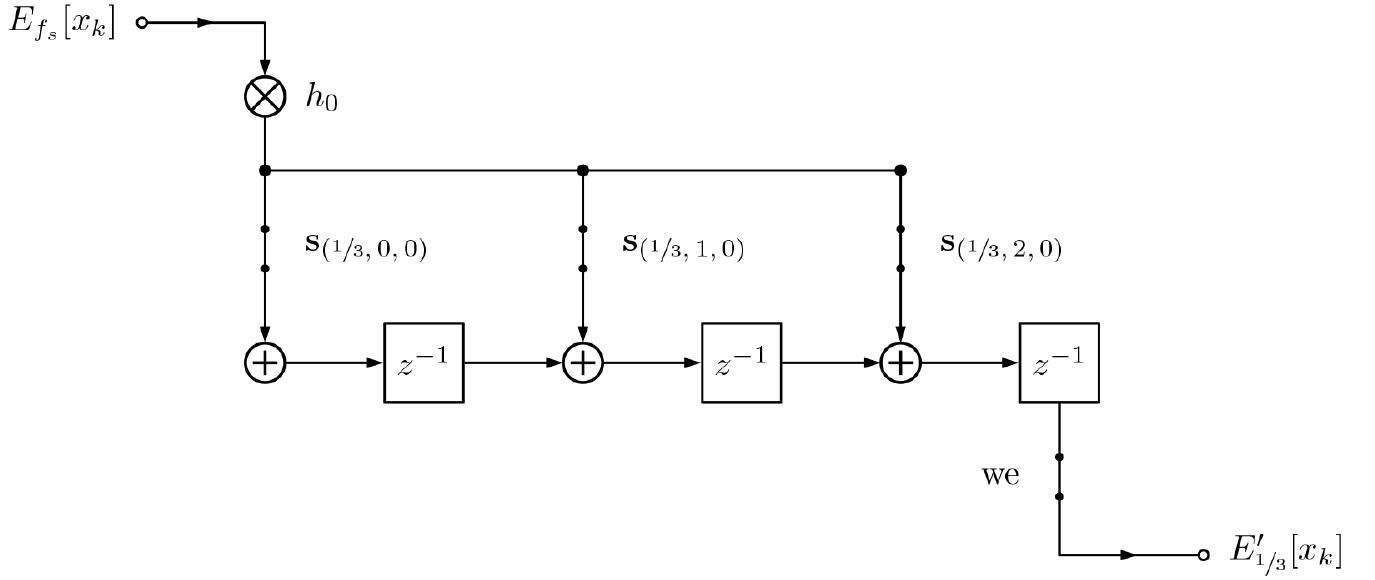}
	\caption[1-D semi-systolic FIR filter for sub-pixel shift $a = \sfrac{1}{3}$]{1-D semi-systolic FIR filter for sub-pixel shift $a = \sfrac{1}{3}$.\label{fig:FIR1}}
\end{figure}
\begin{figure}[ht]
	\includegraphics[width=\linewidth]{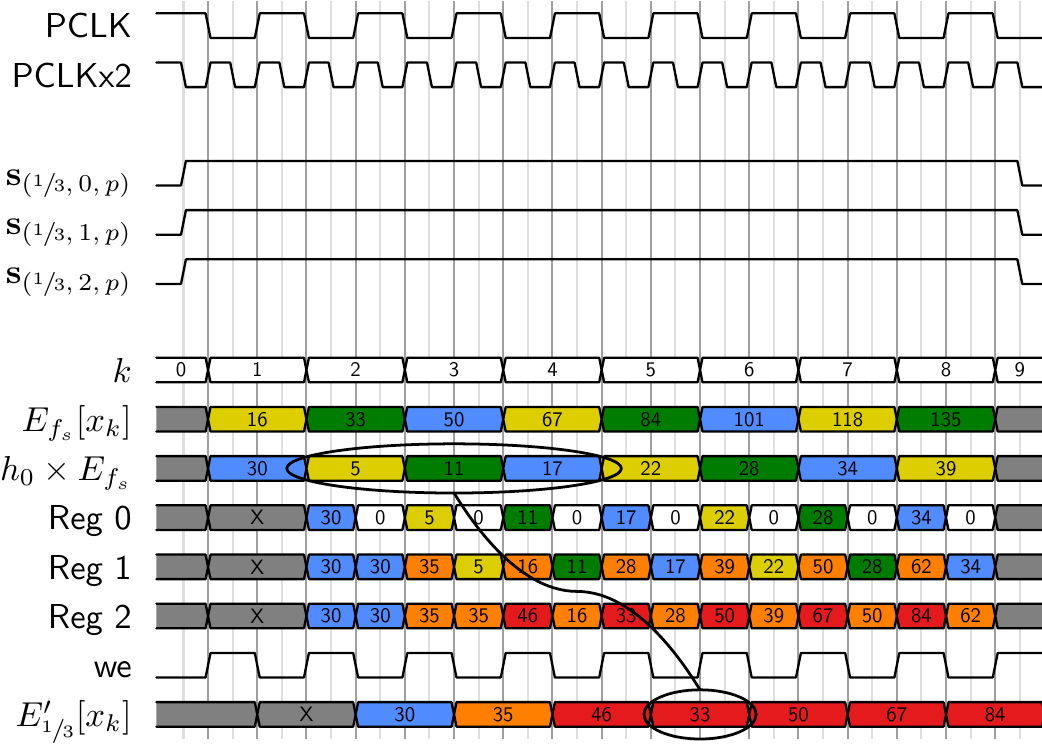}
	\caption[Timing diagram of 1-D semi-systolic FIR filter for $a=\sfrac{1}{3}$]{Timing diagram of FIR filter module with $a=\sfrac{1}{3}$.}
	\label{fig:flowDiagram1}
\end{figure}

\noindent
To refocus with $a=\sfrac{1}{3}$, another FIR filter module is conceived based on Eq.~(\ref{eq:refocusingX:1}) and depicted in Fig.~\ref{fig:FIR1}. 
In reference to the previous FIR filter where $a=\sfrac{0}{3}$, it becomes obvious that the arrangements are identical except for different switch states. The switch state matrix $\vect{s}_{\left(\sfrac{1}{3},\,w,\,p\right)}$ is given by
\begin{align}
\vect{s}_{\left(\sfrac{1}{3},\,w,\,p\right)} =
\begin{bmatrix}
1 & 1 & 1 \\
1 & 1 & 1 \\
1 & 1 & 1 \\
\end{bmatrix}
\, ,
\end{align}
which means that switches remain closed at all times. A corresponding timing diagram is shown in Fig.~\ref{fig:flowDiagram1}. \par
\begin{figure}[ht]
	\centering
	\includegraphics[width=1.0175\linewidth]{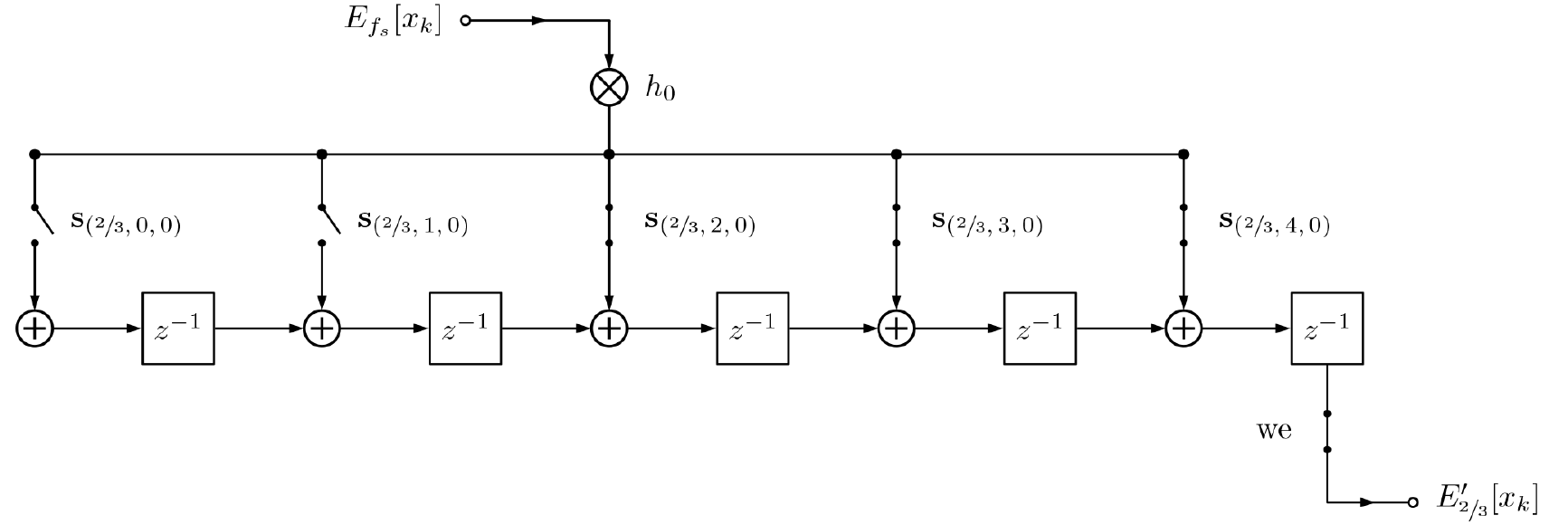}
	\caption[1-D semi-systolic FIR filter for sub-pixel shift $a = \sfrac{2}{3}$]{1-D semi-systolic FIR filter for sub-pixel shift $a = \sfrac{2}{3}$.\label{fig:FIR2}}
\end{figure}

\begin{figure}[H]
	\includegraphics[width=\linewidth]{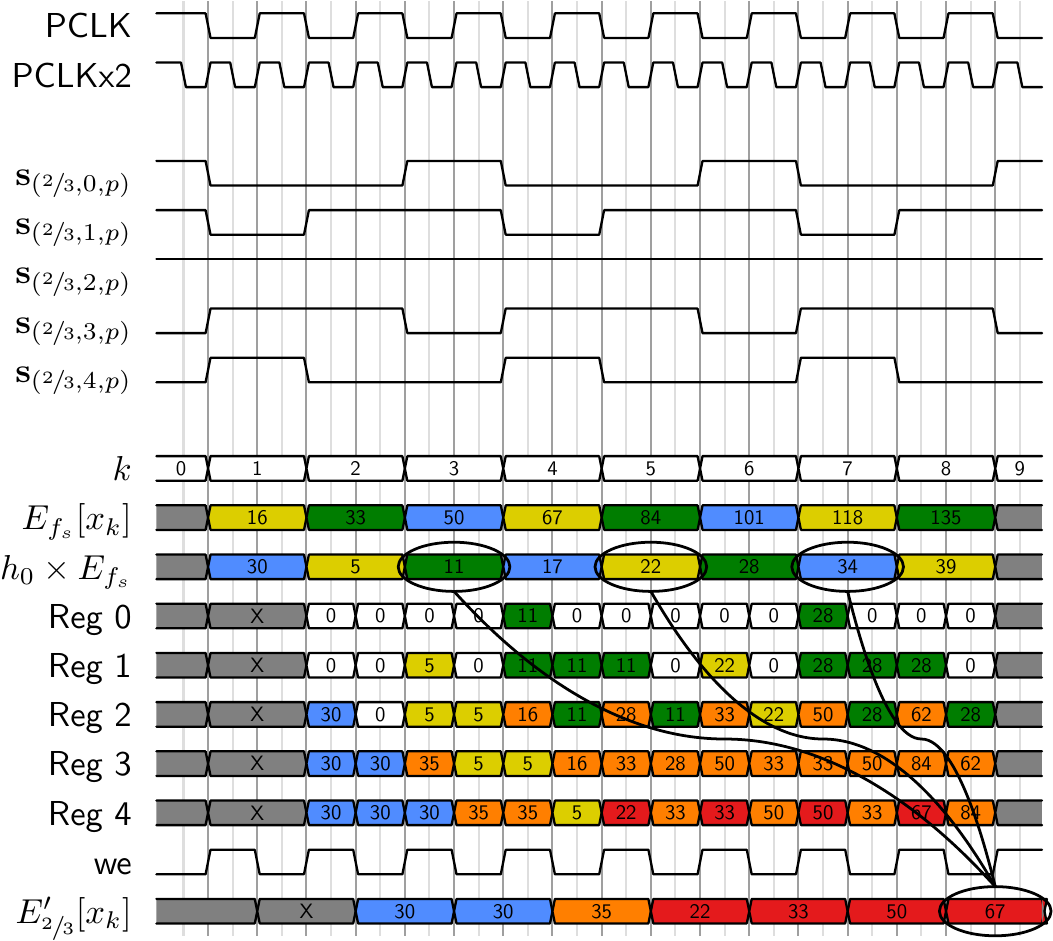}
	\caption[Timing diagram of 1-D semi-systolic FIR filter for $a=\sfrac{2}{3}$]{Timing diagram of FIR filter module with $a=\sfrac{2}{3}$.}
	\label{fig:flowDiagram2}
\end{figure}

\noindent
Figure~\ref{fig:FIR2} depicts an FIR filter according to Eq.~(\ref{eq:refocusingX:2}),
which occupies more PEs due to the fact that the distance between added pixels grows. 
The corresponding switch state matrix $\vect{s}_{\left(\sfrac{2}{3},\,w,\,p\right)}$ is as follows
\begin{align}
\vect{s}_{\left(\sfrac{2}{3},\,w,\,p\right)} =
\begin{bmatrix}
0 & 0 & 1 & 1 & 1 \\
0 & 1 & 1 & 1 & 0 \\
1 & 1 & 1 & 0 & 0 \\
\end{bmatrix}
\, ,
\end{align}
producing a filter behavior shown in Fig.~\ref{fig:flowDiagram2}. \par
\noindent
%
As Fig.~\ref{fig:FIR2} demonstrates,
a large \mbox{1-D} semi-systolic filter module may imply long wires when broadcasting multiplier outputs. Long wires would cause a low-pass filter behavior in the signal transmission, which affects the readability of falling and rising edges and therefore has to be avoided. To keep wires short in the broadcast net, incoming bit words can be distributed to several synchronized latches (buffers) before being merged in adders. %
\noindent
%
\subsection{2-D Module Array}
\label{ssec:verShift}
\begin{figure}[ht]
	\includegraphics[width=\linewidth]{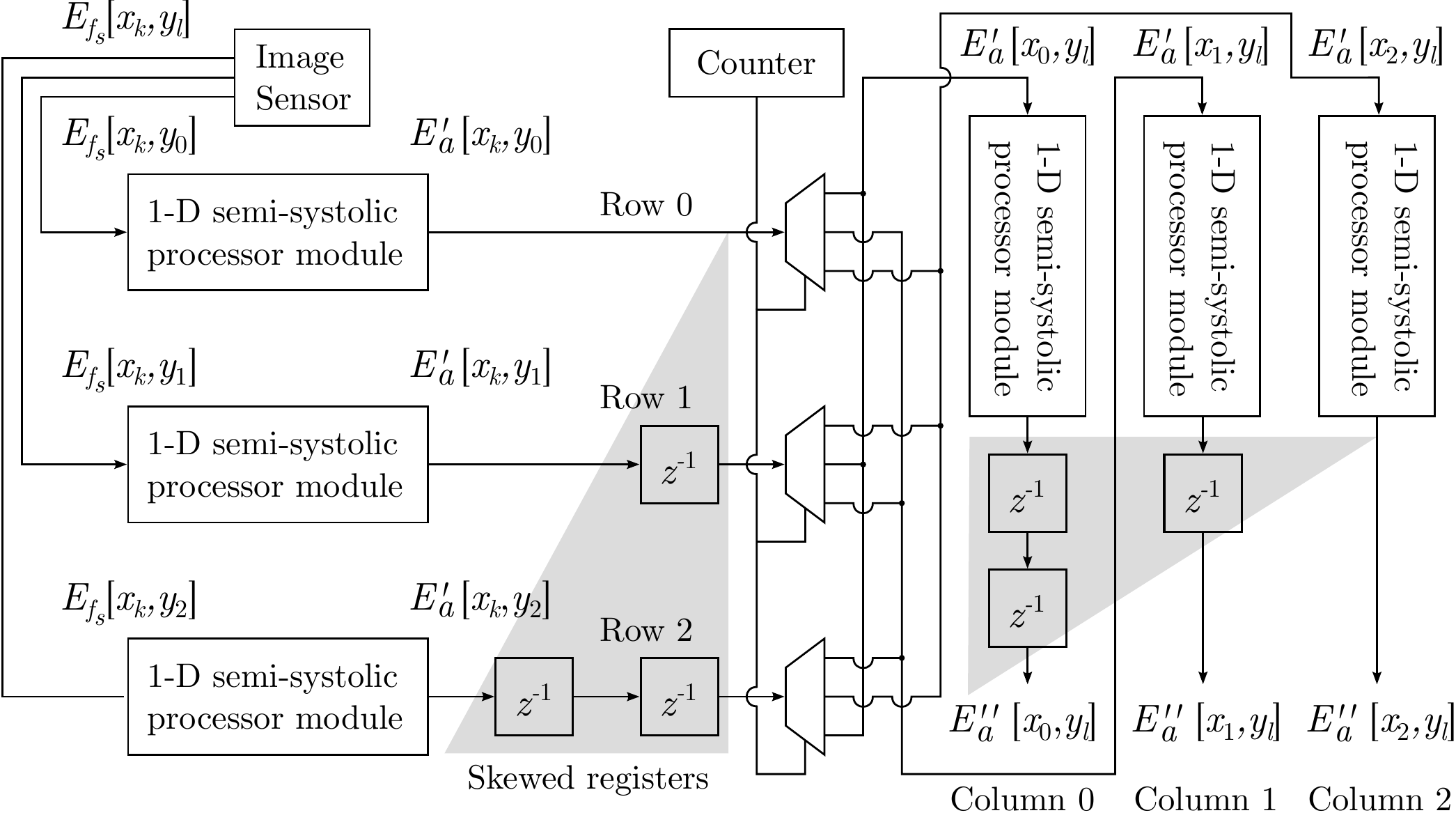}
	\caption[Parallelized 2-D processing module array with $\iota=3$]{Parallelized 2-D processing module array with $\iota=3$.
	}
	\label{fig:verticalShift}
\end{figure}

The proposed FIR filter modules process data in \mbox{1-D} and thus in horizontal or vertical directions only. Fig.~\ref{fig:verticalShift} shows a \mbox{2-D} construct of \mbox{1-D} semi-systolic processor modules to accomplish refocusing by processing data in both dimensions. In this example, the degree of parallelization amounts to $\iota=3$, but could be scaled as desired until limits are reached ($\iota=L$ for image rows, $\iota=K$ for image columns). \par
\noindent
The data flow in Fig.~\ref{fig:verticalShift} is described in the following. First, pixels coming from the sensor are fed into horizontal processor blocks representing semi-systolic FIR filter modules as proposed in the previous section. All semi-systolic processor modules are identical whereas the type relies on the refocusing parameter $a$. In the second stage, horizontally processed data rows $E'_{a}\left[x_k,y_l\right]$ are delayed using skewed registers and assigned to another arrangement of semi-systolic modules making it possible to form an incoming image column $(\text{e.g.}~E'_a\left[x_0,y_l\right])$. 
Here, demultiplexers are driven by a pixel counter to assist in the correct assignment of pixels values. 
This assures that pixels from different rows sharing index $k$ are sent to the same vertical processing unit which produces an image column $(\text{e.g.}~E''_a\left[x_0,y_l\right])$ of the final refocused image. For synchronization purposes, an additional array of skewed registers can be optionally placed behind column processor blocks. \par
\par
%
In order to estimate the computation time, it is assumed hereafter that the hardware system refers to the ideal case of maximum parallelization where $\iota=L$ or $\iota=K$ for each dimension, respectively. Besides, it is supposed that color channels are also parallelized causing no extra time delay. The shift and integration for a single output pixel refocused with $a=\sfrac{1}{M}$ takes $M$ pixel clock cycles in \mbox{1-D} when using twice the pixel clock to process them. Taking this as an example, the overall number of steps $\eta$ to compute a single image $E''_{\sfrac{1}{3}}$ with $K$-by-$L$ resolution is given by
\begin{align}
\eta = 2(\Lambda+M) + 2(K-1) + L-1 \, ,
\end{align}
where $\Lambda$ represents a single clock cycle step to compute the mathematical product of an incoming pixel value. The total computation time $\mathcal{O}$ for a single image can be obtained by
\begin{align}
\mathcal{O}(\eta) = \eta \times T_{\text{PCLK}} \, .
\end{align}
This duration reflects the theoretical time that elapsed from the moment the first pixel $E_{f_s}\left[x_k,y_l\right]$ entered the logic gate until the final output pixel $E''_a\left[x_k,y_l\right]$ is available. When pipelining the data stream, output pixels of a subsequent image arrive directly after that letting the overall computation time for a single frame be represented by the delay time of the computational focusing system. Once the first refocused photograph is received, the number of remaining computational steps $\eta_{\text{sub}}$ for every following image amounts to
\begin{align}
\eta_{\text{sub}} = L-1+K-1 \, .
\end{align}
To assess performance limits of the presented architecture, 
we performed
a benchmark comparison between this approach, the FPGA-based 
implementation of P\'{e}rez~{\it et al.}~\cite{Perez:2014}, 
and a GPU-based approach~\cite{lumsdaine12:_plenoptic_gpus}.
In this comparison, a \mbox{3201-by-3201} pixel image $(K=L=3201)$ with \mbox{291-by-291} micro lenses was computationally refocused in 105.9~ms at 100~MHz clock frequency. Thereby, the micro image resolution is $M=11$ and the output image resolution amounts to \mbox{589-by-589}, which is less than 
$\sfrac{1}{6}$
of the incoming image. Conversely, the proposed \text{semi-systolic} method numerically preserves the incoming spatial resolution by employing an NN-interpolation in $\eta=1+11+3200+1+11+3200+3200$ steps yielding $\mathcal{O}(\eta)=96.2~\mu $s computation time for a single frame when running at 100~MHz pixel clock. Each subsequent frame, however, can be processed in $\eta_{\text{sub}}=3200+3200$ steps which is available every $\mathcal{O}(\eta_{\text{sub}})=64~\mu $s. 
For comparison, an identical implementation based on the GPU implementation by Lumsdaine~{\it et al.}~\cite{lumsdaine12:_plenoptic_gpus} takes approximately 1.38~ms on average, whereas a Matlab implementation
takes approximately 12.1~s per image on average as seen in the overview in Table~\ref{tab:benchmark}. \par
\begin{table}[ht]
	\centering
	\caption{Benchmark of proposed architecture. \label{tab:benchmark}}
	\resizebox{\columnwidth}{!}{%
		\begin{tabular}{|c||c|c|c|c|}
			\hline
			& Proposed design & P\'{e}rez~{\it et al.}~\cite{Perez:2014} & GPU & Matlab \\
			\hline
			Clock frequency & 100 MHz & 100 MHz & 1.35 GHz & 3.40 GHz \\
			Time to compute frame & $96.2~\mu $s & 105.9~ms & 1.38~ms  & 12.1~s \\
			\hline
		\end{tabular}
	}
\end{table}
%
In this comparison, we employed the Spartan-6 XC6SLX45 chip using the ISE WebPACK design software from Xilinx. The refocusing shader were executed on a Fermi architecture GeForce 480M GTX with 2~GB of GDDR5 RAM running at 1200~MHz, connected to a 256~bit bus~\cite{lumsdaine12:_plenoptic_gpus}. For the CPU environment, we used Matlab 7.11.0.584 (R2010b) on an Intel Core i7-3770 CPU @ 3.40~GHz without multi-threading.
\section{Validation}
\label{sec:fpgaimplementation}
\begin{figure}[ht]
	\centering
	\includegraphics[width=\linewidth]{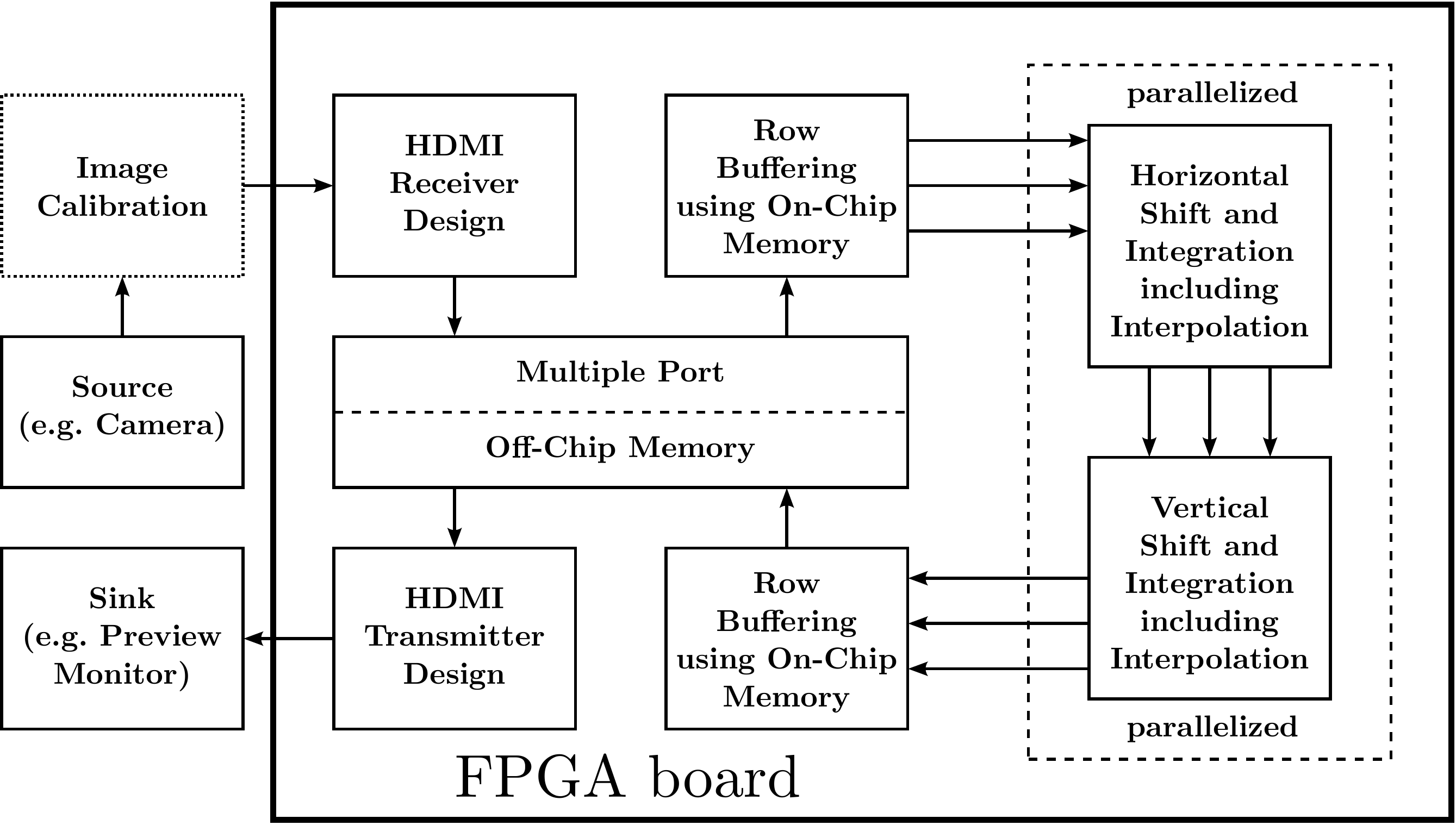}
	\caption[Block diagram for signal processing]{Block diagram (borrowed from \cite{Hahne:14:SPIE}) for experimental validation. Single arrows denote serialized whereas three arrows indicate parallelized data streams. Row buffers are employed to simulate data parallelization in the experiment.}
	\label{fig:blockdiagram}
\end{figure}

%
In this section, we evaluate the functionality of the proposed FPGA-based refocusing hardware design. For that purpose, the \textit{VHSIC Hardware Description Language} (VHDL) is used to configure the FPGA where VHSIC stands for \textit{Very High Speed Integrated Circuit}. A schematic file, generated from a VHDL compiler, is then flashed onto the FPGA chip model XC6SLX45\if which belongs to the Spartan-6 device family\fi. Figure~\ref{fig:blockdiagram} contains a block diagram illustrating the implemented processing architecture used to validate the design proposed in the previous section.
\noindent
The FPGA board features \textit{High-Definition Multimedia Interface} (HDMI) connectors such that video frame transmission is accomplished using the \textit{Transition Minimized Differential Signaling} (TMDS) protocol. TMDS receiver and transmitter designs have been integrated on the FPGA to fulfill deserialization, serialization just as decoding and encoding tasks. Off-chip memory is used for buffering decoded and serialized video frames outside the FPGA since the amount of image data exceeds internal memory storage. %
\par
%
%
In our implementation, a row of switch settings is loaded from a \textit{Look-Up Table} (LUT) every clock cycle starting from the first row again after the last one is reached. 
The switch-state LUTs can be stored in \textit{Block Random-Access Memorys} (BRAMs), which are part of the FPGA. 
%
The integration of multiplier $h_0$ is also achieved using on-chip memory, making it called \textit{stored product}. In accordance with the TMDS protocol specification, a decoded pixel value is of 8 bit depth per color channel, which yields a manageable number of 256 possible results when dividing by $M$. Thus, quotients can be \mbox{pre-calculated} for a specific divisor, $M$, and stored in one BRAM per color channel for each image row. Note that these BRAMs are read-only memories. %
\par
\begin{figure}[ht]
	\hspace*{-.2cm}  
	\resizebox{1.026\columnwidth}{!}{
		\includegraphics[width=\linewidth]{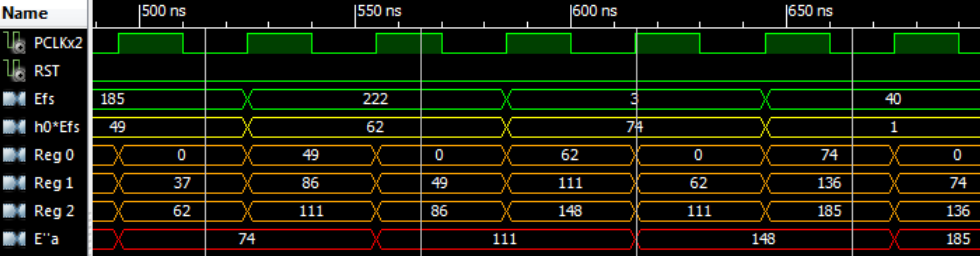}
	}
	\caption[Timing diagram example  from ISE Simulator]{Timing diagram example from ISE Simulator.}
	\label{fig:iSim}
\end{figure}

%
A screenshot from an exemplary timing diagram simulation where $a=\sfrac{1}{3}$ and $T_{\text{PCLK}}=60$~ns is provided in Fig.~\ref{fig:iSim} with the code attached to this article. This VHDL-implemented hardware simulation shows that the filter behaves as expected, justifying the conceived architecture. PCLKx2 can be obtained with a \textit{Phase-Locked Loop} (PLL). \par%
\noindent
An overview of the implemented design comprising a single FIR filter with $a=\sfrac{1}{5}$ is presented in Table~\ref{tab:power} where it can be seen that \textit{Inputs/Outputs} (IOs) and PLLs make up by far most of the power consumption. This is due to the included HDMI transceiver, \textit{Memory Controller Block}~(MCB) and color conversion modules. Parts of these modules may be omitted or replaced by on-board \textit{Integrated Circuits}~(ICs) in a prototyping stage. 
Further, Table~\ref{tab:power} gives indication that adding more FIR filters for full parallalization (maximum $L$ and $K$) is non-critical to power, but may be limited to the number of logic slices in a Spartan-6 device. \par
\begin{table}[ht]
	\centering
	\caption{Utilization Summary for XC6SLX45-CSG324. \label{tab:power}}
	\resizebox{\columnwidth}{!}{%
		\begin{tabular}{
				| c |	
				|S[table-format=3.2]
				|S[table-format=3.0]
				|S[table-format=5.0]
				|S[table-format=2.0]|
			}
			\hline
			{On-chip} & {Power [mW]} & {Used} & {Available} & {Utilization [\%]} \\
			\hline
			Clocks & 82.97 & 8 & {---} & {---} \\
			Logic & 2.68 & 957 & 27288  & 4 \\
			Signals & 12.82 & 1646 & {---} & {---} \\
			IOs & 461.29 & 84 & 218 & 39 \\
			PLLs & 314.69 & 2 & 4 & 50 \\
			MCBs & 189.00 & 1 & 2 & 50 \\
			Quiescent & 79.02 & {---} & {---} & {---} \\
			\hline
			\bfseries{Total} & 1142.46 & {---} & {---} & {---} \\
			\hline
		\end{tabular}
	}
\end{table}
Presented refocusing synthesis formulas require all micro images to be of a consistent size. This is not the case, however, in raw lightfield photographs. As indicated with the experimental architecture in Fig.~\ref{fig:blockdiagram}, micro image cropping remains an external process performed prior to streaming the data to the FPGA. Embedding this process on an FPGA is essential for prototyping, but left for future work. To comply with FIR filter designs in Section~\ref{sec:Shift}, the micro image size is reduced to $M=3$ and $M=5$ for comparison. Lightfield images have been acquired by our custom-built plenoptic camera with an MLA of 281 micro lenses per row and 188 per column. Insightful details on the camera calibration can be found in~\cite{Hahne:Thesis}. \par
\begin{figure}[ht]
	\centering
	\begin{minipage}[b]{.48\linewidth}
		\centering
		\includegraphics[width=\linewidth]{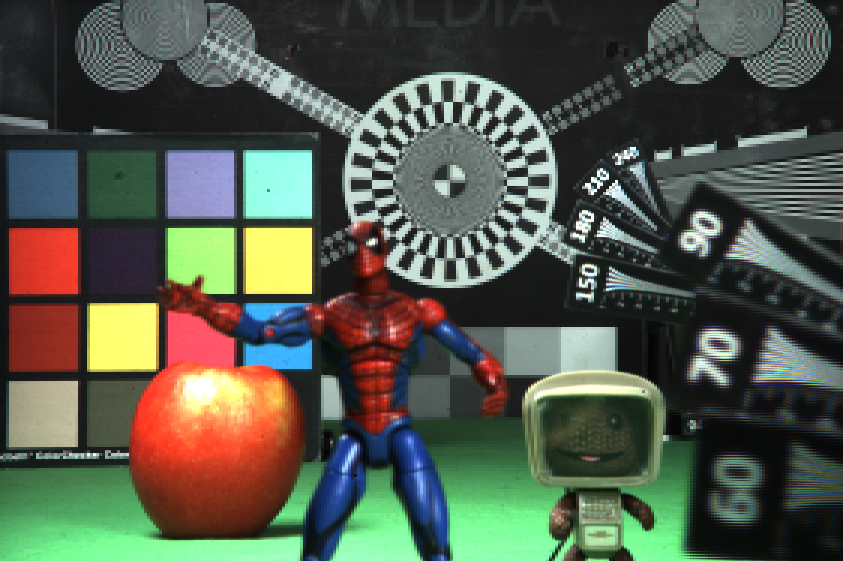}
		\subcaption{$E'_{\sfrac{0}{3}}$\label{fig:FPGArefoc:a}}
	\end{minipage}
	\hfill
	\begin{minipage}[b]{.48\linewidth}
		\centering
		\includegraphics[clip, trim=5px 0 5px 0, width=\linewidth]{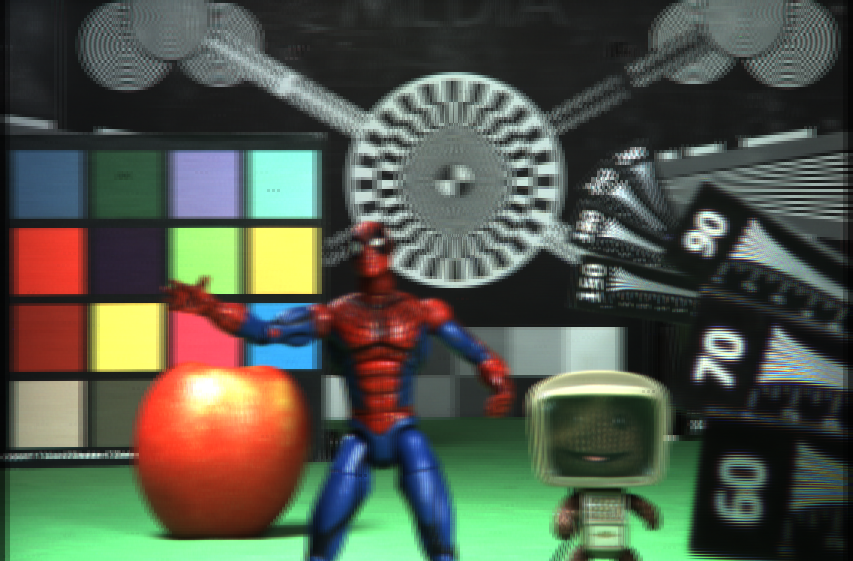}
		\subcaption{$E'_{\sfrac{5}{3}}$\label{fig:FPGArefoc:b}}
	\end{minipage}
	\\
	\vspace{.2cm}
	\begin{minipage}[b]{.48\linewidth}
		\centering
		\includegraphics[width=\linewidth]{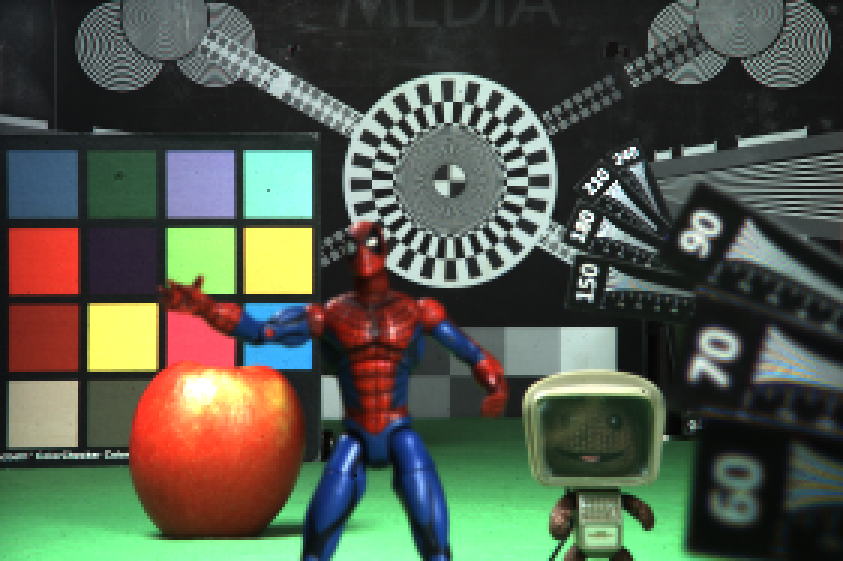}
		\subcaption{$E''_{\sfrac{0}{3}}$\label{fig:FPGArefoc:c}}
	\end{minipage}
	\hfill
	\begin{minipage}[b]{.48\linewidth}
		\centering
		\includegraphics[clip, trim=5px 5px 5px 5px, width=\linewidth]{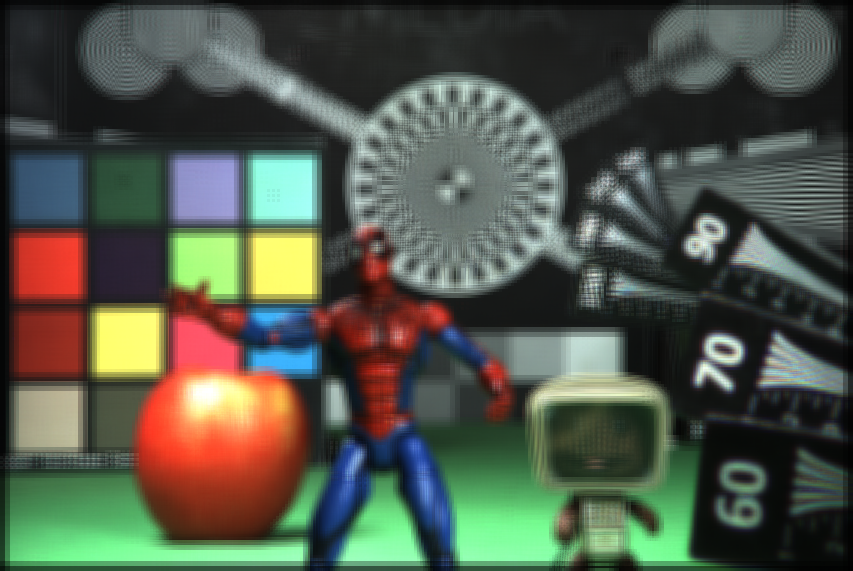}
		\subcaption{$E''_{\sfrac{5}{3}}$\label{fig:FPGArefoc:d}}
	\end{minipage}
	\\
	\vspace{.2cm}
	\begin{minipage}[b]{.48\linewidth}
		\centering
		\includegraphics[width=\linewidth]{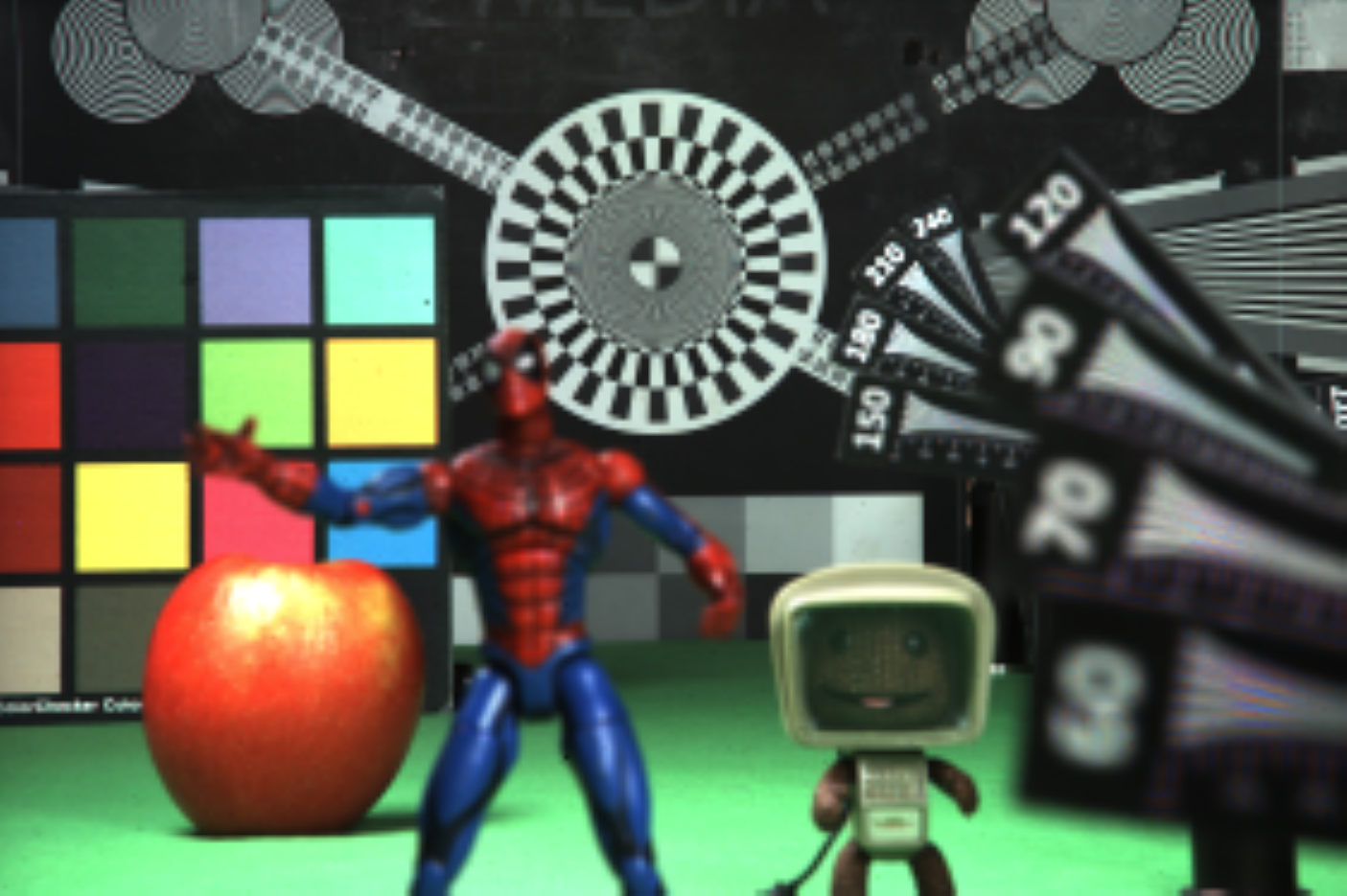}
		\subcaption{$E''_{\sfrac{0}{5}}$\label{fig:FPGArefoc:e}}
	\end{minipage}
	\hfill
	\begin{minipage}[b]{.48\linewidth}
		\centering
		\includegraphics[clip, trim=16px 16px 16px 16px, width=\linewidth]{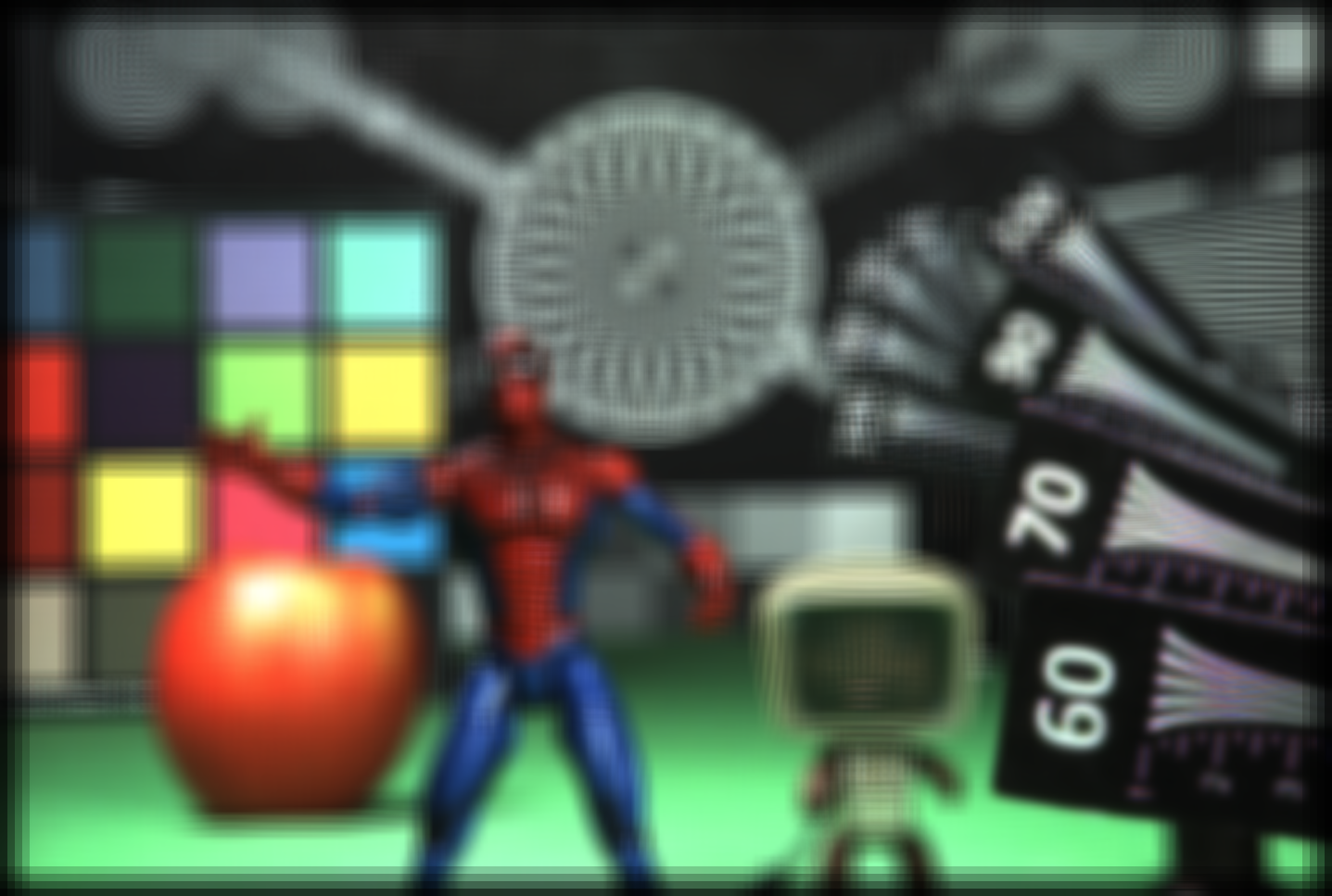}
		\subcaption{$E''_{\sfrac{8}{5}}$\label{fig:FPGArefoc:f}}
	\end{minipage}
	\\
	\caption[Refocused photographs using the proposed architecture]{Refocused photographs using the proposed architecture. Input and output spatial image resolutions amount to \mbox{843-by-561} pixels with $M=3$ in \subref{fig:FPGArefoc:a}~to~\subref{fig:FPGArefoc:d}. Intermediate horizontally processed images are shown in \subref{fig:FPGArefoc:a} and \subref{fig:FPGArefoc:b} whereas \subref{fig:FPGArefoc:c} and \subref{fig:FPGArefoc:d} depict fully refocused images after horizontal and vertical processing with varying $a$. For comparison, output images in \subref{fig:FPGArefoc:e} and \subref{fig:FPGArefoc:f} with \mbox{1405-by-935} pixel resolution expose improved synthetic blur by using a linear interpolation of whole micro images with $M=5$. Reducing a lightfield's angular sampling rate $M$ extends the depth of field~\cite{Hahne:OPEX:16} and leads to blur aliasing in case of angular undersampling~\cite{NG}.}
	\label{fig:FPGArefoc}
\end{figure}

%
Figure~\ref{fig:FPGArefoc} depicts refocused photographs computed by the proposed \mbox{2-D} module array to accomplish {real-time} refocusing. Intermediate results after processing images in a horizontal direction are seen in Figs.~\ref{fig:FPGArefoc:a} and~\ref{fig:FPGArefoc:b}. Their fully refocused counterparts are found in Figs.~\ref{fig:FPGArefoc:c} and~\ref{fig:FPGArefoc:d}. Closer inspection of \ref{fig:FPGArefoc:d} indicates aliasing in blurred regions. This is due to an undersampled directional domain as there are only \mbox{3-by-3} samples per micro image ($M=3$) in the incoming lightfield capture. Aliasing in synthetic image blur is an observation Ng already pointed out in his thesis~\cite{NG}. To combat the aliasing problem, the author suggests to sufficiently increase the micro image sampling rate $M$. Figs.~\ref{fig:FPGArefoc:e} and~\ref{fig:FPGArefoc:f} show refocused images obtained from a raw capture with a native micro image resolution of \mbox{5-by-5} pixels ($M=5$) using a linear interpolation instead of nearest neighbor. There, it can be seen that aliasing artifacts are satisfyingly suppressed.
%
\noindent
%
%
\begin{figure}[ht]
	\begin{minipage}[t]{\linewidth}
		\centering
		\begin{minipage}[t]{.31\linewidth}
			\centering
			\includegraphics[scale=.2357142857]{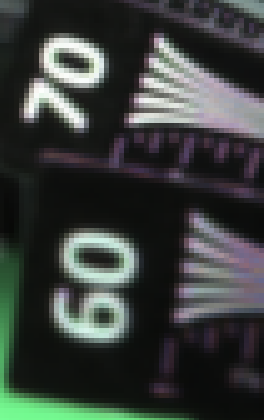}
			\subcaption{NN interp. $E''_{\sfrac{5}{5}}$ \\ (while refocusing)\label{fig:resEnhance:b}}
		\end{minipage}
		\begin{minipage}[t]{.31\linewidth}
			\centering
			\includegraphics[scale=.2357142857]{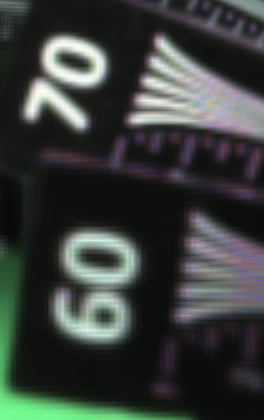}
			\subcaption{NN interp. $E''_{\sfrac{4}{5}}$ (while refocusing)\label{fig:resEnhance:f}}
		\end{minipage}
		\begin{minipage}[t]{.31\linewidth}
			\centering
			\includegraphics[scale=.2357142857]{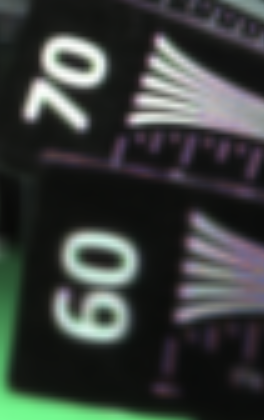}
			\subcaption{Lin. interp. $E''_{\sfrac{5}{5}}$ \\ (while refocusing)\label{fig:resEnhance:d}}
		\end{minipage}
		\\
		\vspace{.3cm}
	\end{minipage}
	\begin{minipage}[t]{\linewidth}
		\centering
		\begin{minipage}[t]{.31\linewidth}
			\centering
			\includegraphics[scale=.2357142857]{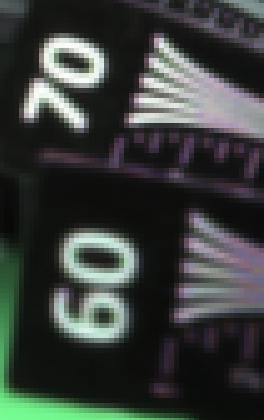}
			\subcaption{NN interp. $E''_{\sfrac{5}{5}}$ \\ (after refocusing)\label{fig:resEnhance:c}}
		\end{minipage}
		\begin{minipage}[t]{.31\linewidth}
			\centering
			\includegraphics[scale=.2357142857]{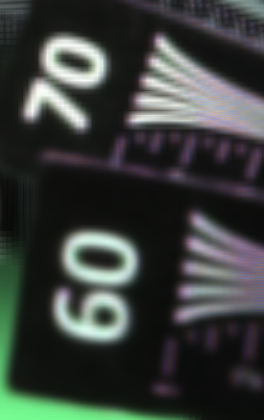}
			\subcaption{NN interp. $E''_{\sfrac{6}{5}}$ \\ (while refocusing)\label{fig:resEnhance:g}}
		\end{minipage}
		\begin{minipage}[t]{.31\linewidth}
			\centering
			\includegraphics[scale=.2357142857]{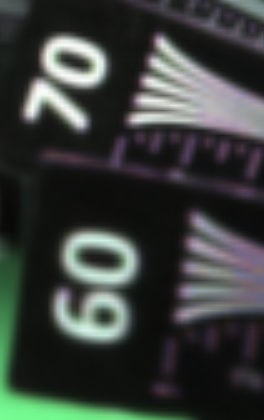}
			\subcaption{Lin. interp. $E''_{\sfrac{5}{5}}$ \\ (after refocusing)\label{fig:resEnhance:e}}
		\end{minipage}
		\\
		\vspace{.3cm}
	\end{minipage}
	\caption[Resolution comparison]{Resolution comparison where~\subref{fig:resEnhance:b}, \subref{fig:resEnhance:c}, \subref{fig:resEnhance:d} and \subref{fig:resEnhance:e} show the same region refocused with $a=\sfrac{5}{5}$ using different interpolation techniques during and after shift and integration. Images in \subref{fig:resEnhance:f} and \subref{fig:resEnhance:g} are NN-interpolated versions with varying $a$ indicating significant variation of the spatial resolution when compared to \subref{fig:resEnhance:b} and \subref{fig:resEnhance:c}. Effective resolution is more consistent when using linear interpolation (e.g. compare \subref{fig:resEnhance:c}, \subref{fig:resEnhance:g} and \subref{fig:resEnhance:e}).
		\label{fig:resEnhance}}
\end{figure}\noindent
%
\par%
A comparison of output image resolutions using the inherent NN-interpolation of proposed FIR filters is provided in Fig.~\ref{fig:resEnhance}. Results in sub-figures \ref{fig:resEnhance:b}~to~\ref{fig:resEnhance:e} suggest that interpolating micro images while refocusing with $a \in \mathbb{Z}$ using Eq.~(\ref{eq:refocusingX:3}) corresponds to a conventional \mbox{2-D} image interpolation. On the contrary, an effective resolution enhancement can be observed when comparing Fig.~\ref{fig:resEnhance:b} where $a=\sfrac{5}{5}$ with Fig.~\ref{fig:resEnhance:f} where $a=\sfrac{4}{5}$, which are both computed from the same raw image using NN-interpolation. Given that respective objects are acceptably well covered by their depth of field and exhibit best focus, it is possible to state that improved resolution is obtained by refocusing with non-integer numbers ($a \centernot\in \mathbb{Z}$). This effective resolution variation is a consequence of the micro image repetition and the interleaving filter kernel for the refocusing synthesis yielding identical values for adjacent output pixels when $a \in \mathbb{Z}$, but varying intensities for contiguous pixels if $a \in \mathbb{R}$. This can be seen by inspecting output data streams $E'_a[x_k]$ of the timing diagrams in Fig.~\ref{fig:flowDiagram0} and Fig.~\ref{fig:flowDiagram1}. To work towards consistency in spatial resolutions for varying $a$, it is thus essential to employ linear interpolation prior to distributing micro image pixels through the FIR broadcast net. 
A positive side effect in upsampling micro images is that refocused image slices  $E''_{a}\left[x_k, y_l\right]$ are not only interpolated in spatial-domain, but also sub-sampled along depth as demonstrated in~\cite{Hahne:OPEX:16}. 
%
\section{Conclusion}
\label{sec:discussion}
%
This paper has demonstrated methods to derive optimized FIR refocusing filter kernels for a time-~and cost-efficient hardware implementation. Simulating the conceived architecture proved that real-time refocusing can be accomplished with a computation time of $96.24~\mu$s per frame reducing the delay time \if interval\fi by 99.91~\% in comparison to a previous state-of-the-art attempt. By interpolating micro images, it was shown how to retain the numerical sensor resolution in refocused photographs. %
The proposed architecture can serve as a groundwork for \textit{Application-Specific Integrated Circuit} (ASIC) chips. \par
A limitation of the results is that timing delays have been simulated and need to be verified using chip analyzing tools. As the number of required PEs grows with higher image resolutions, it may exceed the gate count capacity of the FPGA in full parallelization. Besides this, care needs to be taken to prevent long wires in the broadcast net. For the hardware system's reliability, it is also recommended to convert semi-systolic arrays into a full-systolic architecture. 
To achieve consistency in micro image size ($M$), cropping of the same has to be integrated as a preceding processing stage on the FPGA chip. Furthermore, a bilinear interpolation ought to be implemented to replace micro image repetition (NN-interpolation) and work towards consistent effective resolutions in refocused images, although this will cause additional delays. \par%
A competitive design approach may conceive a refocusing architecture based on the \textit{Fourier Slice Photography Theorem}. It is, however, expected that the Fourier transform produces larger time delays. A considerable alternative to an FPGA-based implementation is the employment of a GPU as this takes less design effort, however, by inducing larger delays and more power consumption. \par%
Deployment of proposed design to an FPC is thought to be impractical, since there is a fundamental difference between SPC and FPC with regards to the optical design (number of micro lenses and focus position of MLA). On the algorithmic level, SPC refocusing is a pixel-based integration whereas an FPC requires the integration of overlapping areas of shifted micro image patches such that a refocusing algorithm has to be designed specific to the type of plenoptic camera.
%
\bibliographystyle{Bibliography/IEEEtranTIE}
\bibliography{Bibliography/memoirthesis} 
{	
	\begin{IEEEbiography}[{\includegraphics[width=1in,height=1.25in,clip,keepaspectratio]{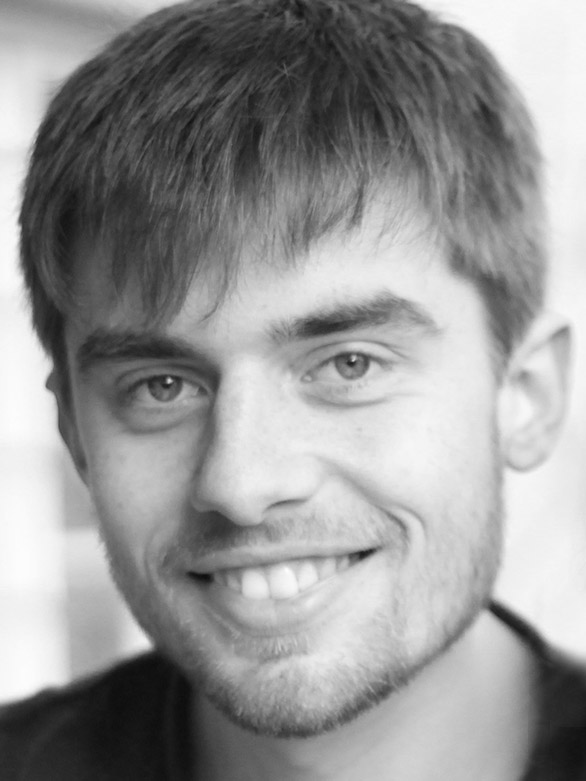}}]
		{Christopher Hahne} is affiliated with BASF subsidiary trinamiX GmbH, Ludwigshafen, Germany, where he currently works as the Manager of Simulation \& Software on adaptive 3-D sensing. He received the BSc degree in Media Technology from the University of Applied Sciences, Hamburg, Germany, in 2012 whilst working at R\&D departments of Rohde~\&~Schwarz GmbH~\&~Co. KG, Munich, Germany in 2010 and Arnold~\&~Richter Cinetechnik GmbH~\&~Co. KG, Munich, Germany in 2011. Subsequently, he became a visiting student at Brunel University, London, UK, in 2012 and was awarded the doctoral degree in Computer Science from the University of Bedfordshire, UK, in 2016 in a bursary-funded PhD programme. \\ \\
	\end{IEEEbiography}
	
	\begin{IEEEbiography}[{\includegraphics[width=1in,height=1.25in,clip,keepaspectratio]{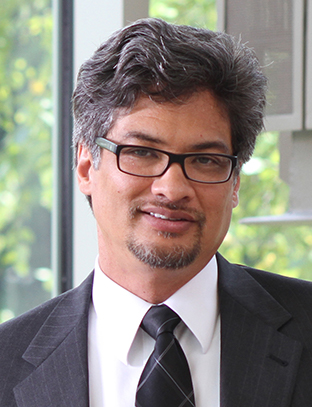}}]
		{Andrew Lumsdaine} (SM'15) received  He is an internationally recognized expert in the area of high-performance computing who has made important contributions in many of the constitutive areas of HPC. In particular, he has contributed in the areas of HPC systems, programming languages, software libraries, and performance modeling. His work in HPC has been motivated by data-driven problems (large-scale graph analytics), as well as more traditional computational science problems. In addition, outside of the realm of HPC, he has done seminal work in the area of computational photography and plenoptic cameras. In his career, he has published more than 200 articles in top journals and conferences and holds 15 patents. Dr. Lumsdaine also has contributed important software artifacts to the research community, especially in the area of Message Passing Interface (MPI). He also is active in a number of standardization efforts with important contributions to the MPI specification, the C++ programming language, and the Graph 500. \\ \\ 
	\end{IEEEbiography}
	
	\begin{IEEEbiography}[{\includegraphics[width=1in,height=1.25in,clip,keepaspectratio]{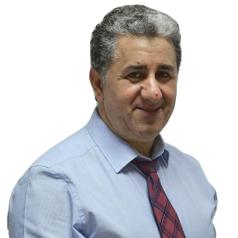}}]
		{Amar Aggoun} is currently the Head of School of Mathematics and Computer Science and Professor of Visual Computing at the University of Wolverhampton. He received the "Ingenieur d'\'etat" degree in Electronics Engineering in 1986 from Ecole Nationale Polytechnique d'Alger (Algiers, Algeria) and a PhD degree in Electronic Engineering from the University of Nottingham. His academic carrier started at the University of Nottingham where he held the positions of research fellow in low power DSP architectures and visiting lecturer in Electronic Engineering and Mathematics. In 1993 he joined De Montfort University as a lecturer and progressed to the position of Principal Lecturer in 2000. In 2005 he joined Brunel University as Reader in Information and Communication Technologies. From 2013 to 2016, he was at the University of Bedfordshire as Head of School of Computer Science and Technology. He was also the director of the Institute for Research in Applicable Computing which oversees all the research within the School. His research is mainly focused on 3D Imaging and Immersive Technologies and he successfully secured and delivered research contracts worth in excess of \pounds6.9M, funded by the research councils UK, Innovate UK, the European commission and industry. Amongst the successful project, he was the initiator and the principal coordinator and manager of a project sponsored by the EU-FP7 ICT-4-1.5-Networked Media and 3D Internet, namely Live Immerse Video-Audio Interactive Multimedia (3D VIVANT). He holds three filed patents, published more than 200 peer-reviewed journals and conference publications and contributed to two white papers for the European Commission on the Future Internet. He also served as Associate Editor for the IEEE/OSA Journal of Display Technologies.
	\end{IEEEbiography}

	\begin{IEEEbiography}[{\includegraphics[width=1in,height=1.25in,clip,keepaspectratio]{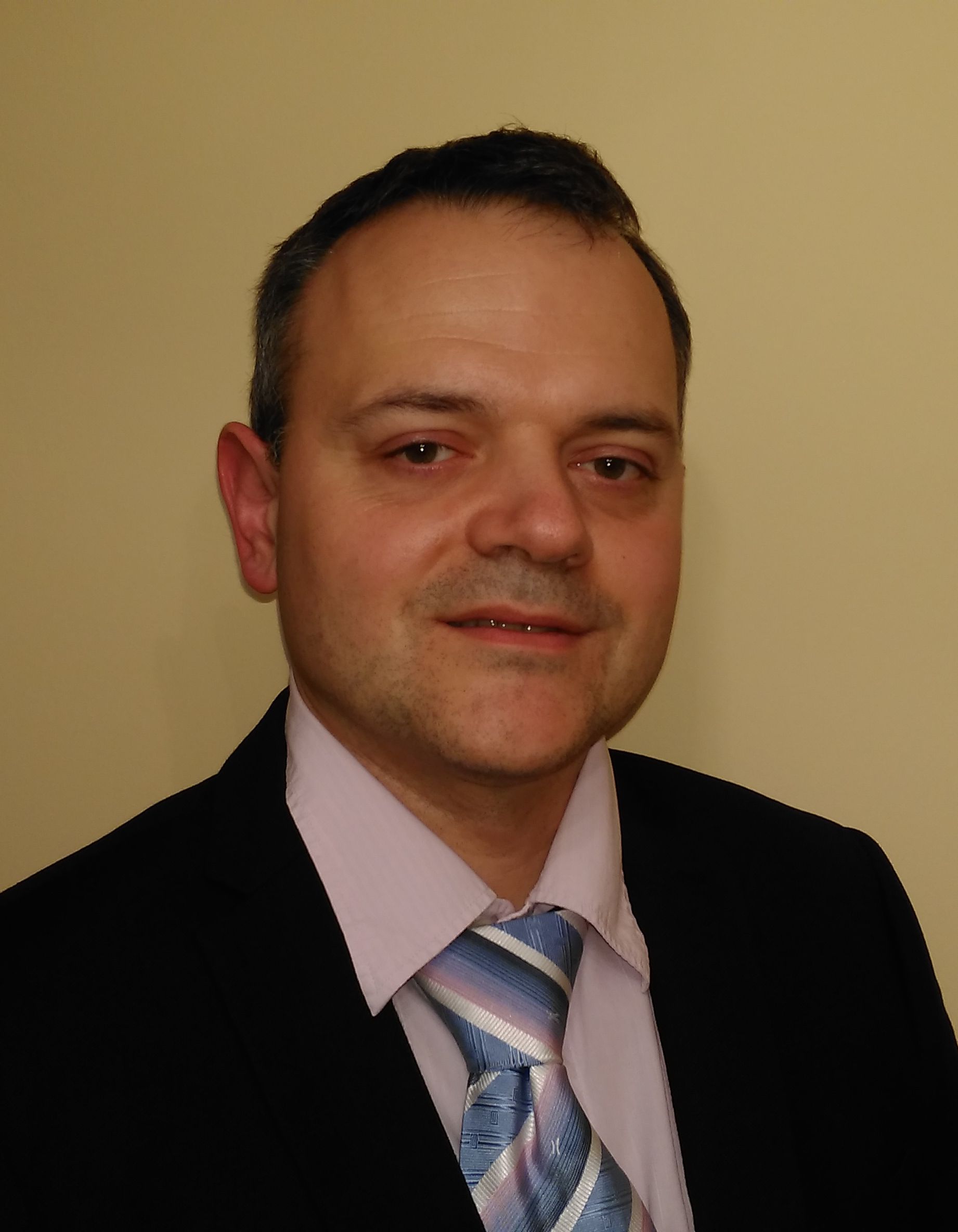}}]
	{Vladan Velisavljevic} (M'06, SM'12) is Reader (Associate Professor) in Visual Systems Engineering at the School of Computer Science and Technology and Head of the Centre for Research in Signals, Sensors and Wireless Technology, University of Bedfordshire, UK, since 2011. Previously, he was Senior Research Scientist at Deutsche Telekom Laboratories, University of Technology Berlin, Germany, in 2006-2011 and Doctoral Assistant at LCAV, EPFL, Switzerland, in 2001-2005. He got PhD from EPFL in 2005 in the field of signal and image processing. Vladan has published more than 60 peer-reviewed journal and conference publications and 2 book chapters. He serves as Associate Editor for Elsevier Signal Processing: Image Communication and for IET Journal of Engineering and he is co-chair of the IEEE ComSoc MMTC Interest Group on 3D Processing and Communications. He was General Chair of the IEEE MMSP 2017, Lead Guest Editor for special issue on Visual Signal Processing for Wireless Networks at the IEEE Journal of Selected Topics in Signal Processing in February 2015 and special session organizer at 3DTV-Con 2015 and IEEE ICIP 2011. He was also Associate Editor for IEEE ComSoc MMTC R-Letters and Member of the Review Board for the IEEE ComSoc Multimedia Communications TC. He has served as a TPC member and reviewer for a number of conferences and journals. \\ \\ 
	\end{IEEEbiography}

\end{document}